\documentclass[twocolumn,twocolappendix]{aastex631}

\shorttitle{Pa$\alpha$ Star Forming Sequence}
\shortauthors{Neufeld et al.}

\begin{document}

\title{FRESCO: The Paschen-$\alpha$ Star Forming Sequence at Cosmic Noon}

\author[0000-0002-6558-9894]{Chloe Neufeld}
\affiliation{Astronomy Department, Yale University, 52 Hillhouse Ave, New Haven, CT 06511, USA}

\author[0000-0002-8282-9888]{Pieter van Dokkum}
\affiliation{Astronomy Department, Yale University, 52 Hillhouse Ave, New Haven, CT 06511, USA}

\author[0000-0002-8320-2198]{Yasmeen Asali}
\affiliation{Astronomy Department, Yale University, 52 Hillhouse Ave, New Haven, CT 06511, USA}

\author[0000-0002-9672-3005]{Alba Covelo-Paz}
\affiliation{Department of Astronomy, University of Geneva, Chemin Pegasi 51, 1290 Versoix, Switzerland}

\author[0000-0001-6755-1315]{Joel Leja}
\affiliation{Department of Astronomy \& Astrophysics, The Pennsylvania State University, University Park, PA 16802, USA}
\affiliation{Institute for Computational \& Data Sciences, The Pennsylvania State University, University Park, PA 16802, USA}
\affiliation{Institute for Gravitation and the Cosmos, The Pennsylvania State University, University Park, PA 16802, USA}

\author[0000-0002-3101-8348]{Jamie Lin}
\affiliation{Department of Physics and Astronomy, Tufts University, 574 Boston Avenue, Medford, MA 02155, USA}

\author[0000-0003-2871-127X]{Jorryt Matthee}
\affiliation{Institute of Science and Technology Austria (ISTA), Am Campus 1, 3400 Klosterneuburg, Austria}

\author[0000-0001-5851-6649]{Pascal A. Oesch}
\affiliation{Department of Astronomy, University of Geneva, Chemin Pegasi 51, 1290 Versoix, Switzerland}
\affiliation{Cosmic Dawn Center (DAWN), Niels Bohr Institute, University of Copenhagen, Jagtvej 128, K\o benhavn N, DK-2200, Denmark}

\author[0000-0001-9687-4973]{Naveen A. Reddy}
\affiliation{Department of Physics and Astronomy, University of California, Riverside, 900 University Avenue, Riverside, CA 92521, USA}

\author[0000-0003-4702-7561]{Irene Shivaei}
\affiliation{Centro de Astrobiolog\'{i}a (CAB), CSIC-INTA, Carretera de Ajalvir km 4, Torrej\'{o}n de Ardoz, 28850, Madrid, Spain}

\author[0000-0001-7160-3632]{Katherine E. Whitaker}
\affiliation{Department of Astronomy, University of Massachusetts, Amherst, MA 01003, USA}
\affiliation{Cosmic Dawn Center (DAWN), Niels Bohr Institute, University of Copenhagen, Jagtvej 128, København N, DK-2200, Denmark}

\author[0000-0003-3735-1931]{Stijn Wuyts}
\affiliation{Department of Physics, University of Bath, Claverton Down, Bath, BA2 7AY, UK}

\author[0000-0003-2680-005X]{Gabriel Brammer}
\affiliation{Cosmic Dawn Center (DAWN), Niels Bohr Institute, University of Copenhagen, Jagtvej 128, K\o benhavn N, DK-2200, Denmark}

\author[0000-0001-9002-3502]{Danilo Marchesini}
\affiliation{Department of Physics and Astronomy, Tufts University, 574 Boston Avenue, Medford, MA 02155, USA}

\author[0000-0003-0695-4414]{Michael V. Maseda}
\affiliation{Department of Astronomy, University of Wisconsin-Madison, 475 N. Charter St., Madison, WI 53706 USA}

\author[0000-0003-3997-5705]{Rohan P. Naidu}\altaffiliation{NHFP Hubble Fellow}
\affiliation{MIT Kavli Institute for Astrophysics and Space Research, 77 Massachusetts Ave., Cambridge, MA 02139, USA}

\author[0000-0002-7524-374X]{Erica J. Nelson}
\affiliation{Department for Astrophysical and Planetary Science, University of Colorado, Boulder, CO 80309, USA}

\author[0000-0001-7512-1606]{Anna Velichko}
\affiliation{Department of Astronomy, University of Geneva, Chemin Pegasi 51, 1290 Versoix, Switzerland}
\affiliation{Institute of Astronomy of V. N. Karazin Kharkiv National University, Svobody square 4, 61022 Kharkiv, Ukraine}

\author[0000-0001-8928-4465]{Andrea Weibel}
\affiliation{Department of Astronomy, Université de Genève, Chemin Pegasi 51, 1290 Versoix, Switzerland}

\author[0000-0003-1207-5344]{Mengyuan Xiao}
\affiliation{Department of Astronomy, University of Geneva, Chemin Pegasi 51, 1290 Versoix, 
Switzerland}

\correspondingauthor{Chloe Neufeld}
\email{chloe.neufeld@yale.edu}

\begin{abstract}

We present results from the \textit{JWST} First Reionization Epoch Spectroscopically Complete Observations survey (FRESCO) on the star forming sequence of galaxies at $1.0<z<1.7$, around the peak of the cosmic star formation history. Star formation rates (SFRs) are measured from the redshifted,  {relatively} dust-insensitive Paschen-$\alpha$ emission line, and stellar mass measurements include the F444W (4.4 $\mu$m; rest-frame H) band. We find SFRs of galaxies with  {log($M_*/M_\odot)>9.5$} that are lower than found in many earlier studies by up to 0.6 dex, but in good agreement with recent results obtained with the Prospector fitting framework. The difference log(SFR(Pa$\alpha$)-SFR(\textsc{Prospector})) is -0.09 $\pm$ 0.04 dex at $10^{10-11} M_\odot$. We also measure the empirical relation between Paschen-$\alpha$ luminosity and rest-frame H band magnitude and find that the scatter is only 0.04 dex lower than that of the SFR-M* relation and is much lower than the systematic differences among relations in the literature due to various methods of converting observed measurements to physical properties. We additionally identify examples of sources -- that, with standard cutoffs via the UVJ diagram, would be deemed quiescent -- with significant  {(log(sSFR)$>-11$ yr$^{-1}$)}, typically extended, Paschen-$\alpha$ emission. Our results may be indicative of the potential unification of methods used to derive the star forming sequence with careful selection of star forming galaxies and independent star formation rate and stellar mass indicators.

\end{abstract}

\keywords{Galaxy evolution (594) --- Galaxy properties (615) --- Star formation (1569) --- Scaling relations (2031)}


\section{Introduction} \label{sec:intro}

The star forming sequence (SFS) of galaxies --- the relation between star formation rate (SFR) and stellar mass (M*) --- is a useful probe of galaxy evolution, with implications for how galaxies build up their mass and eventually quench (e.g., \citealp{noeske2007,elbaz2011,speagle2014,whitaker2014,lee2015,renzini2015,salmon2015,schreiber2015,tomczak2016,santini2017,leslie2020,leja2022,merida2023,popesso2023}). The relation has been measured in both observations and simulations in a large number of previous studies (see \citealp{speagle2014} and \citealp{popesso2023} for reviews). 

Generally, the relation is  {found to have very low scatter of 0.2-0.3 dex in the $10^{9}M_\odot<M*<10^{10}M_\odot$ mass regime \citep{noeske2007,schreiber2015}}. This tight relation indicates that star formation histories (SFHs) are smooth: stellar mass grows steadily with SFR at a consistent rate across these galaxies  {\citep{noeske2007,daddi2007,renzini2009,finlator2011,whitaker2014,salmon2015}.}  {There are conflicting results on the dependence of the scatter on stellar mass: higher scatter in the relation has been observed at the low mass end \citep{atek2022,merida2023}, as well as an increase in scatter as mass increases (\citealp{santini2017,boogaard2018,sherman2021}). }

The normalization of the SFS increases with redshift, indicating that gas accretion occurred at a higher rate earlier in the universe (\citealp{tomczak2016}), and the peak star formation for individual galaxies with high masses is predicted to occur earlier than that of lower masses \citep{thomas2005}. Previous studies find that the sequence can be well fit by a broken power law with a turnover that evolves with redshift; the slope below this turnover mass is $\sim 1$, while above it is shallower, with suppressed specific SFR (e.g., \citealp{whitaker2014,schreiber2015,tomczak2016,leja2022}). This has several physical implications: \cite{popesso2023} suggest that the time-dependence of the turnover mass is due to black hole feedback and the availability of cold gas in the haloes. Looking at the bulge-to-total mass ratios, \cite{Cooke2023} find evidence that the presence of this turnover mass is due to the evolution of the specific star formation rate of the disk rather than bulge growth (see also \citealp{abramson2014,whitaker2015}).

Key questions surrounding the relation remain. While there is general agreement in a large number of studies about the scatter and the existence of this tight relation, there are conflicting results about the exact quantification of the normalization and slope of the SFS among observational studies. \cite{speagle2014} find that these conflicts can mostly be ascribed to different methods used to infer the relation (i.e., inferring SFRs and stellar masses), including selection effects, assumed IMFs, dust extinction, and SED fitting methods, and those issues have not been fully resolved in the years since. The normalization of the SFS determined from galaxy formation simulations is also 0.2-0.5 dex lower than that found in observational studies \citep{somervilladave2015,tomczak2016,popesso2023}. This was only recently mitigated with the use of updated masses and SFRs from the \textsc{Prospector} Bayesian inference fitting framework and spatially resolved star formation measurements  \citep{leja2019,nelson2021,leja2022}. These differences lead to the question of what the ``true" relation is, and which method best recovers it.

Many studies use the observed bimodality between the rest-frame colors of galaxies to select for star-forming galaxies, but this is highly dependent on the methodology of color measurements due to spatially varying stellar populations \citep{kriek2010,whitaker2012b,belli2015}. A related method is to fit spectral energy distribution (SED) models and use a threshold on the specific SFRs (sSFRs) to select quiescent galaxies \citep{merlin2018}, but lower sSFRs have associated uncertainties due to lower intrinsic brightness and can be contaminated by emission from AGN or dust heating by older stellar populations. Moreover, \cite{eales2017} and \cite{feldman2017} argue that the distribution of galaxies is actually unimodal, with no observable separation between star-forming and quenched galaxies. Ambiguity between star-forming and quenched galaxies can lead to artificially flattening the slope of the SFS at high masses \citep{renzini2015,shivaei2015}, and \cite{leja2022} measure differences of up to 0.5 dex in the normalization of the SFS when using several standard pre-selection methods. 

Methods for inferring star formation rates and stellar masses can also affect the measured SFS. For example, some studies measure the SFR using the UV stellar continuum, which is ideal for high redshift observations where the rest-UV emission is redshifted into optical wavelengths. This provides a measure of the emission coming from massive young stars, which thus traces recent star formation \citep{kennicutt2012}. However, this method is sensitive to interstellar dust attenuation, only probing starlight that is unabsorbed by dust. A majority of star formation is obscured by dust, particularly in massive galaxies; attempts to mitigate this are very sensitive to the assumed attenuation laws \citep{salim2020}. Other studies use IR emission in conjunction with UV emission (e.g., \citealp{whitaker2014,lee2015,tomczak2016,shivaei2017,popesso2023}), thus obtaining a combination of the unabsorbed UV light emitted by young, massive stars and their re-radiated IR emission from dust. This too can have associated uncertainties, as IR emission from dust heating can be contaminated by asymptotic giant branch (AGB) stars in old stellar populations and active galactic nuclei. Similarly, various studies look to the full distribution of flux across different wavelengths using SED fitting to infer SFRs (e.g., \citealp{salmon2015,santini2017,leja2022}). While this is the most self-consistent approach, it is difficult to assess uncertainties:  {mass and SFR may be correlated which would lead to an underestimate of the scatter in the SFR-stellar mass relation, and some studies have found that these properties in individual galaxies may be anticorrelated due to uncertainties in dust attenuation \citep{salmon2015,Kurczynski2016,curtislake2021}.}

Some studies have found that measurements of the stellar mass and SFR for an individual galaxy can be anticorrelated due to uncertainties in the dust attenuation

Another commonly used SFR indicator is nebular emission from hydrogen recombination: recently formed massive stars ionize their surrounding gas, and hydrogen recombination cascades produce emission lines \citep{kennicut1998}. Unlike other indicators such as UV emission, nebular line SFRs are much less sensitive to the assumed star formation history. In the optical regime, previous studies have used H$\alpha$ or H$\beta$ (e.g., \citealp{whitaker2014,renzini2015,shivaei2015}). A common practice to correct for optical dust attenuation is to measure the H$\alpha$/H$\beta$ Balmer decrement. However, besides being a difficult measurement, as H$\beta$ is at least $\sim3$ times fainter than H$\alpha$, this correction misses any emission that is not present in either line due to being fully obscured or optically thick in these lines. Other studies have used longer wavelength hydrogen recombination lines that are less sensitive to dust, such as the Paschen series (e.g., \citealp{rieke2009,cleri2022,gimenez2022,reddy2023}), or the Brackett series (e.g., \citealp{pasha2020}). Using JWST NIRSpec observations, \cite{reddy2023} find that SFRs obtained using Paschen-$\alpha$ emission can be larger than those obtained using Balmer lines by 25\%, as expected if some of the star formation is optically thick.

The SFS is especially important to study at cosmic noon ($z\sim1-3$), when the cosmic star formation density was at its peak \citep{madau2014}. While rest-frame NIR hydrogen recombination lines provide obvious benefits over their optical counterparts for clean SFR determinations, they have until recently only been accessible from the ground for the nearest galaxies, as emission beyond $\sim$2.4$\mu$m is inaccessible (e.g., Paschen-$\alpha$ becomes inaccessible at $z>0.280$). With \textit{JWST}, we can now observe longer wavelength SFR tracers at cosmic noon. In this work, we use Pa$\alpha$ (1.875 $\mu$m), the n=4$\rightarrow$3 hydrogen recombination line, as a SFR indicator to probe the relation between SFR and stellar mass in galaxies at redshifts $z\sim1.4$. Using \textit{JWST} NIRCam/grism in the 4.4 $\mu$m band, we are able to measure SFRs using Pa$\alpha$ emission lines, with stellar masses measured in the rest-frame NIR, which allows us to measure the SFR-M* relation using a relatively simple and independently derived method. 

We use Pa$\alpha$ emission line data obtained using JWST NIRCam/grism observations taken as part of the FRESCO survey, as described in Section \ref{sec:data}. In Section \ref{sec:deriving}, we discuss the methods in which we derive physical properties of galaxies in our sample, including obtaining stellar masses using rest-frame NIR magnitude and SFR from Pa$\alpha$ emission, and in Section \ref{sec:sfs}, we show the resulting relation between SFR for our sample. We discuss our results in the context of previous studies in Section \ref{sec:discussion} and provide a summary in Section \ref{sec:conclusion}.

Throughout the paper, we use a \cite{chabrier2003} IMF for stellar mass and SFR calculations and give all magnitudes in the AB system \citep{okegunn}. We assume a $\Lambda$CDM cosmology with $\Omega_M=0.3$, $\Omega_\lambda=0.7$, and a Hubble constant $H_0=70$ km s$^{-1}$ Mpc$^{-1}$.

\section{Data}\label{sec:data}

\subsection{FRESCO Survey}\label{sec:fresco}
We use F444W NIRCam/grism spectra and imaging data in the GOODS North and South fields, obtained between November 2022 and February 2023 from the First Reionization Epoch Spectroscopically Complete Observations survey (FRESCO; Cycle 1 GO-1895; PI Oesch, \citealp{oesch23}), a \textit{JWST} cycle 1 53.8 hour medium program. This survey covers an area of $\sim 60$ arcmin$^2$ in each GOODS field with two 2x4 NIRCam/grism mosaics. All the {\it JWST} data used in this paper can be found in MAST: \dataset[10.17909/gdyc-7g80]{http://dx.doi.org/10.17909/gdyc-7g80}.

The FRESCO survey used grismR for a single dispersion direction over the field, with spectra for most galaxies covering a maximal wavelength range of 3.8 to 5.0 $\mu$m. The spectroscopic exposure time per pointing was 7 ks, allowing for a 5$\sigma$ line sensitivity of $2\times10^{-18}$ erg s$^{-1}$ cm$^{-2}$ at a resolution of $R\sim1600$ for compact sources, translating to Paschen-$\alpha$ luminosities of $L_{\text{PaA}}=10^{40.57}$ erg s$^{-1}$ at redshift $z\sim1.7$. Grism observations provide a complete sample of all emission line sources in the field and reliable spectroscopic redshifts, as well as emission line maps that can be useful for determining where exactly emission is coming from in a galaxy. The wavelength coverage enables line measurements and resolved maps of Pa$\alpha$ at $z\sim$ 1.4. In addition to the F444W filter, the FRESCO survey includes medium band images using the F182M and F210M filters. The exposure time per pointing for direct imaging was 0.9 ks, and the three filters (F182M, F210M, and F444W) have 5$\sigma$ depths (measured in 0.32" diameter circular apertures) of 28.4, 28.2, and 28.2 mag, respectively, as shown in \cite{oesch23}. 

The NIRCam slitless grism data were reduced using the \textsc{grizli} software\footnote{\url{https://github.com/gbrammer/grizli}}\citep{grizli} on the publicly available NIRCam grism configuration files\footnote{\url{https://s3.amazonaws.com/grizli-v2/JwstMosaics/v7/index.html}}. A detailed description of the data reduction process and the derivation of photometric catalogs will be provided in Brammer et al. (in prep). To summarize, raw files are obtained from the MAST archive, aligned to a Gaia-matched reference frame, and each exposure is aligned to the direct images for a given visit. Following \cite{kashino2023}, we obtain continuum-subtracted spectra by using a running median filter along each row with a 12 pixel central gap to subtract continuum emission and minimize self-subtraction. Detection images for each field were obtained using a stack of F210M and F444W FRESCO imaging. The GOODS fields have also been observed in several other surveys, including 3D-HST\footnote{\url{https://archive.stsci.edu/prepds/3d-hst/}} \citep{brammer2012,skelton2014,momcheva2016}, a near-infrared spectroscopic survey using the \textit{Hubble Space Telescope}. Multi-wavelength photometric catalogs were derived by running SExtractor \citep{sextractor} on HST and JWST images, ranging from $\sim$0.4-5.0 $\mu$m, PSF-matched to the F444W filter.  {Fluxes were measured in circular apertures of radii 0.08", 0.16", 0.25" and 0.35" and scaled to the fluxes measured in Kron-like apertures \citep{kron1980} by SExtractor in the F444W image, then corrected to total flux based on the fraction of encircled flux in each Kron ellipse on the F444W PSF. This correction factor allows for the derivation of stellar mass from total flux, but leaves colors as those measured in the circular apertures. All fluxes were also corrected for Milky Way foreground extinction using the extinction model from \cite{fitzpatrickmassa2007}. Source segmentation maps tracing the extended morphology of sources in the F444W imaging were then used to extract 1D spectra and emission line fluxes. }

\subsection{Sample Selection}\label{sec:sample}
\begin{figure}
    \centering
    \includegraphics[width=\columnwidth]{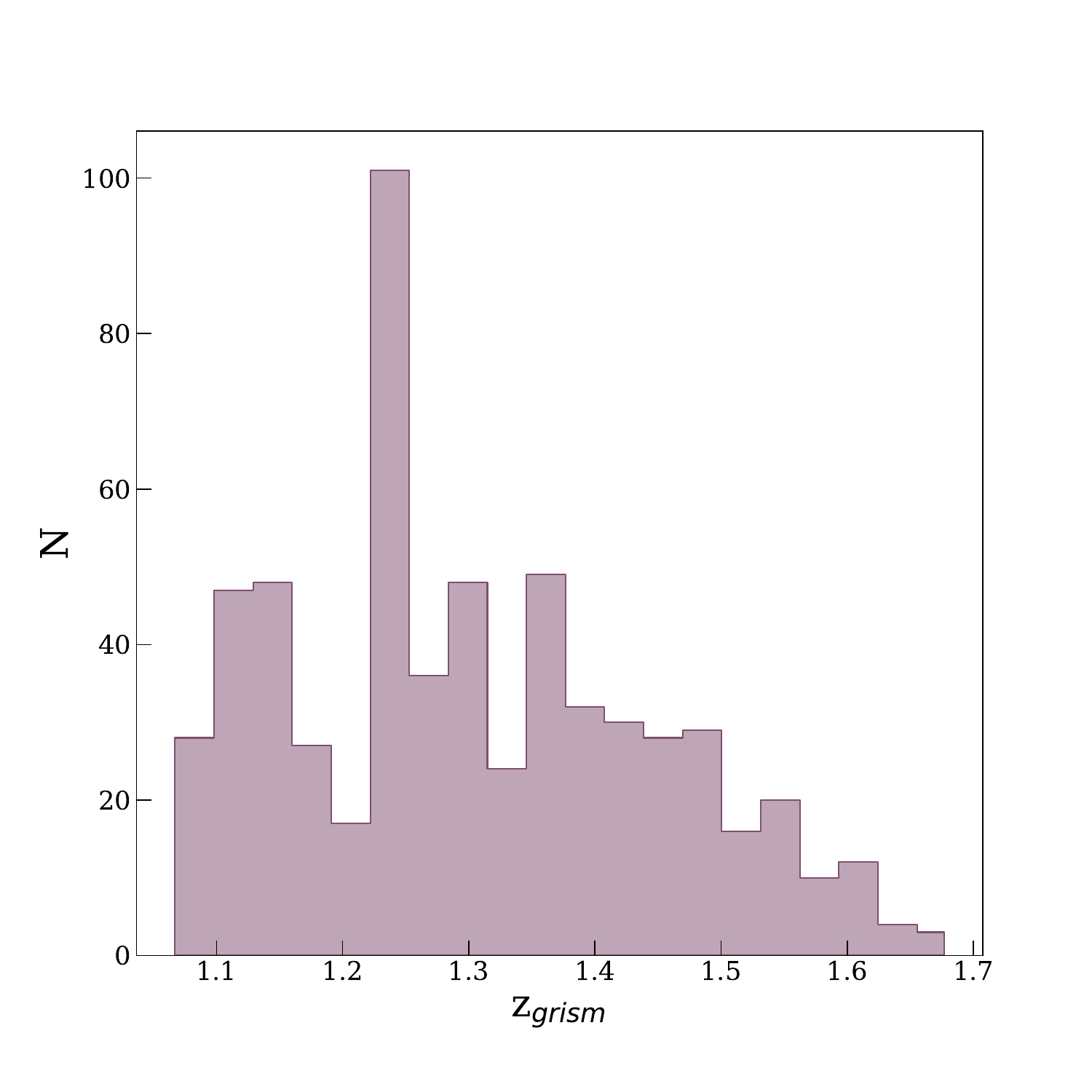}
    \caption{Distribution of spectroscopic redshifts for our Paschen-$\alpha$ sample, which probes $1.0<z<1.7$, with a majority of sources around $z\sim1.2-1.4$.}
    \label{fig:fig1}
\end{figure}

Sources with Paschen-$\alpha$ S/N $>$ 3 are  {originally selected for visual inspection} from the grizli extracted catalog. Subsequently, two individuals visually inspect each source using SpecVizitor software\footnote{\url{https://github.com/ivkram/specvizitor}} and categorize them based on the quality of the line fit and redshift probability distribution into five categories: 1) contaminated by bright sources in the field, 2) absence of detectable lines, 3) uncertain about the line fit quality, 4) robust fit with a single line and consistent morphological shape between the line and imaging (F444W), and 5) robust fit with multiple lines. The individual inspection catalogs are then merged to create the final catalog. Sources flagged as robust (quality flags 4 and 5) by both or only one of the inspectors are categorized as robust and semi-robust, respectively. Where needed, we computed a weighting based on the flag distribution of each inspector and down-weighted those flagged by the more lenient inspector. In this study we adopt sources flagged as robust ($\sim$30\% of the original sample) and semi-robust ($\sim$40\% of the original sample) in the final catalog, as well as further cuts of Paschen-$\alpha$ SNR $>$ 5 and  {95\% confidence width} $z_{\rm{width}}<$ 0.005\footnote{ {95\% confidence width of the redshift fit is defined by $z_{\rm{width}} = (z_{975} - z_{025})/ (1+z_{50}) / 2$, where $z_{975}$ and $z_{025}$ are the upper and lower 95 percentile redshifts and $z_{50}$ is the median redshift of the fits.}}, resulting in a sample of 609 galaxies  {(82\% robust and 18\% semi-robust)} with a redshift distribution shown in Figure \ref{fig:fig1}.

We use EA$z$Y \citep{eazy} SED fitting to estimate physical properties in the following section, such as rest-frame colors and stellar masses, from the flux corrected 0.7" diameter aperture photometry, with redshifts set to those inferred from the grism spectrum of each galaxy. We adopt a 14 template set\footnote{\url{https://github.com/gbrammer/eazy-photoz/tree/master/templates/sfhz}} with redshift-dependent SFHs that can be fit as nonnegative linear combinations to allow for flexibility (see, e.g., \citealp{kokorev2022,gould2023}). 

In Appendix \ref{appendix:mass} we compare our default EA$z$Y masses to those of other codes, in particular \textsc{Bagpipes} \citep{carnall2018} and \textsc{Prospector} \citep{johnson2021}. We use EA$z$Y as the default for several reasons. With F444W (rest-frame NIR) photometric measurements we should constrain the dominant mass component of these galaxies, making a simple model favorable. EA$z$Y is one of the simplest SED fitting methods available, and our results are thus easily reproducible. Several previous studies also use similar EA$z$Y template sets to obtain stellar masses (see, e.g., \cite{sherman2020} for a discussion on the reliability of EA$z$Y for obtaining stellar mass estimates). 

\section{Deriving the Star Forming Sequence}\label{sec:deriving}
The FRESCO survey is well-suited for observing the SFS at cosmic noon. At redshifts of $z\sim1-2$, the 4.4 micron band targets the rest-frame near infrared (NIR) emission of galaxies. This filter is thus especially suited for tracing the dominant stellar mass component - low-mass stars, which typically make up most of the total luminosity in the NIR (e.g., \citealp{wen2013,nagaraj2021}). Additionally, Paschen-$\alpha$ ($\lambda=1.8751\mu$m) falls in the rest-frame NIR, so we can observe this  {relatively} dust-insensitive SFR tracer as well. We are then able to independently observe the relation between SFR (derived from Pa$\alpha$) and $M^*$ (derived using SED fitting that incorporates the rest-frame NIR) at cosmic noon.

\subsection{Rest-frame Colors}\label{sec:rfcolors}
We differentiate between quenched and star forming galaxies in our sample using the rest-frame U-V and V-J colors via the UVJ diagnostic, which is a standard method of separating star forming and quenched systems into distinct regions while breaking the degeneracy between quiescence and dust content (e.g., \citealp{labbe2005,wuyts2007,williams2009,bundy2010,cardamone2010,whitaker2011,brammer2011,patel2012}). 
The EA$z$Y SED fits include interpolated rest-frame fluxes in the U, V, and J bands, from which we derive rest-frame U-V and V-J colors. We use the selection criteria for quiescent galaxies in the redshift range for our sample ($z\sim1-2$) as described in \cite{muzzin2013} and \cite{mowla2019}:

\begin{equation}
    \begin{tabular}{c}
        $U-V>1.3$ \\
        $V-J<1.5$  \\
        $U-V>0.88(V-J)+0.59$ 
    \end{tabular}
\end{equation}

\begin{figure}
    \centering
    \includegraphics[width=\columnwidth]{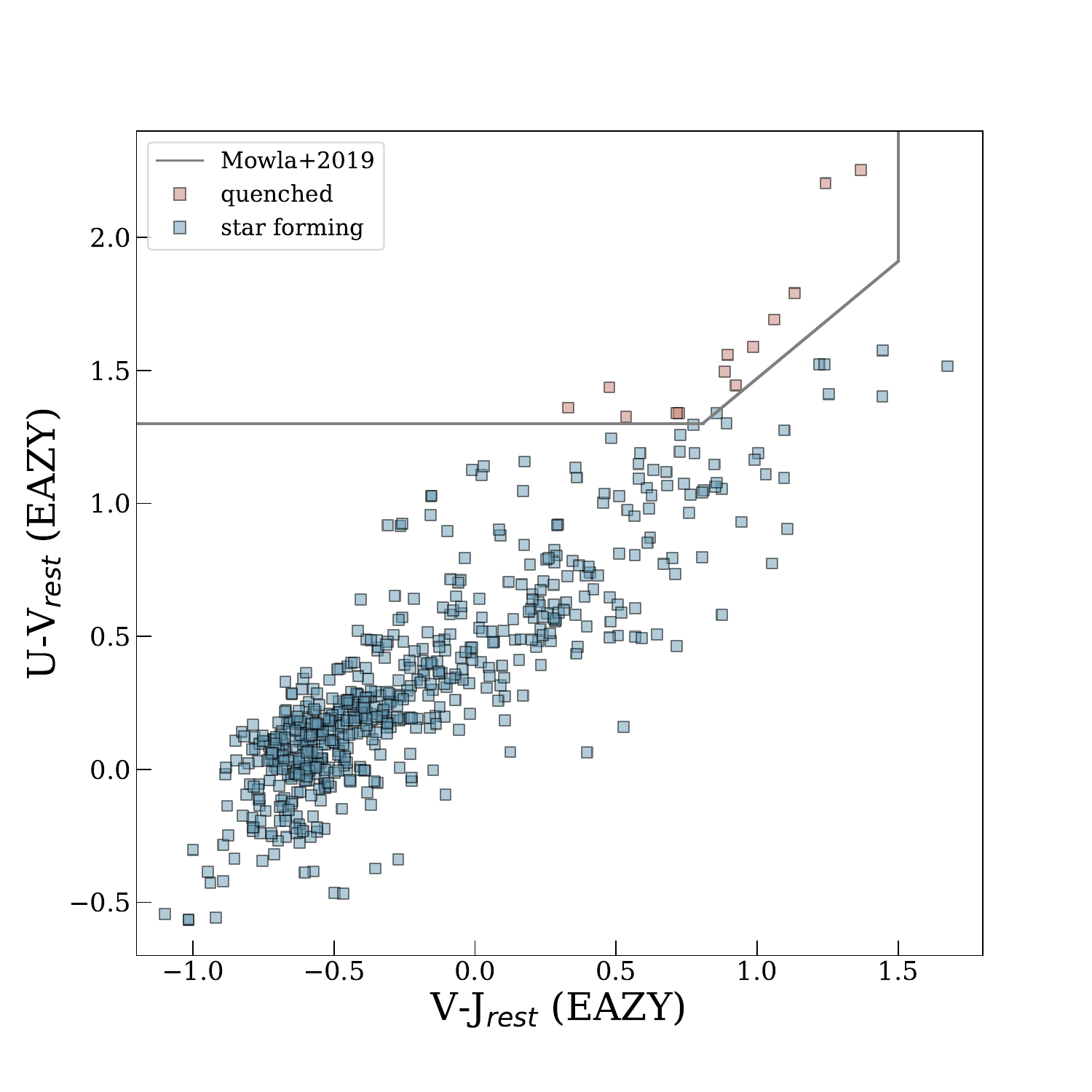}
    \caption{SED fitting derived rest-frame U-V vs. V-J colors (the UVJ diagram) for the 609 galaxies in our Paschen-$\alpha$ selected sample. We use the \cite{mowla2019} criteria to separate quenched (red) vs. star forming (blue) galaxies. Note that there is a population of quenched galaxies in this emission-line-selected sample.}
    \label{fig:fig2}
\end{figure}

From this, we determine which galaxies are classically defined as ``quenched'' or ``star forming,'' with the two regimes and the boundaries separating them shown in Figure \ref{fig:fig2}. 13 galaxies - $\sim2$\% of our sample - would be classified as quenched via the UVJ diagram despite being selected for having  {significant (log(sSFR)$>$-11 yr$^{-1}$, \citealp{fontana2009})} Paschen-$\alpha$ emission lines and thus active star formation. We discuss the caveats of using this method to select star-forming galaxies in the context of spatially resolved emission line maps in Section \ref{sec:quenched}.

We also use our spectral data and the SED fits to measure fluxes in the rest-frame H band ($1.6 \mu$m); this NIR band is directly observed for our sample of $z=1-1.7$ galaxies.

\subsection{EAzY Rest-frame NIR Magnitudes and Stellar Masses}\label{sec:massnir}
As noted aboved, including rest-frame NIR bands in SED fitting should lead to more reliable masses, as we incorporate the dominant stellar component in SED fitting rather than relying on extrapolations using other bands. Previous studies have observed a correlation between absolute magnitude in NIR bands and stellar mass at $z=0$ (e.g., \citealp{wen2013,nagaraj2021}), and here we can explore this relation at $z\sim1.4$. 

\begin{figure}
    \centering
    \includegraphics[width=\columnwidth]{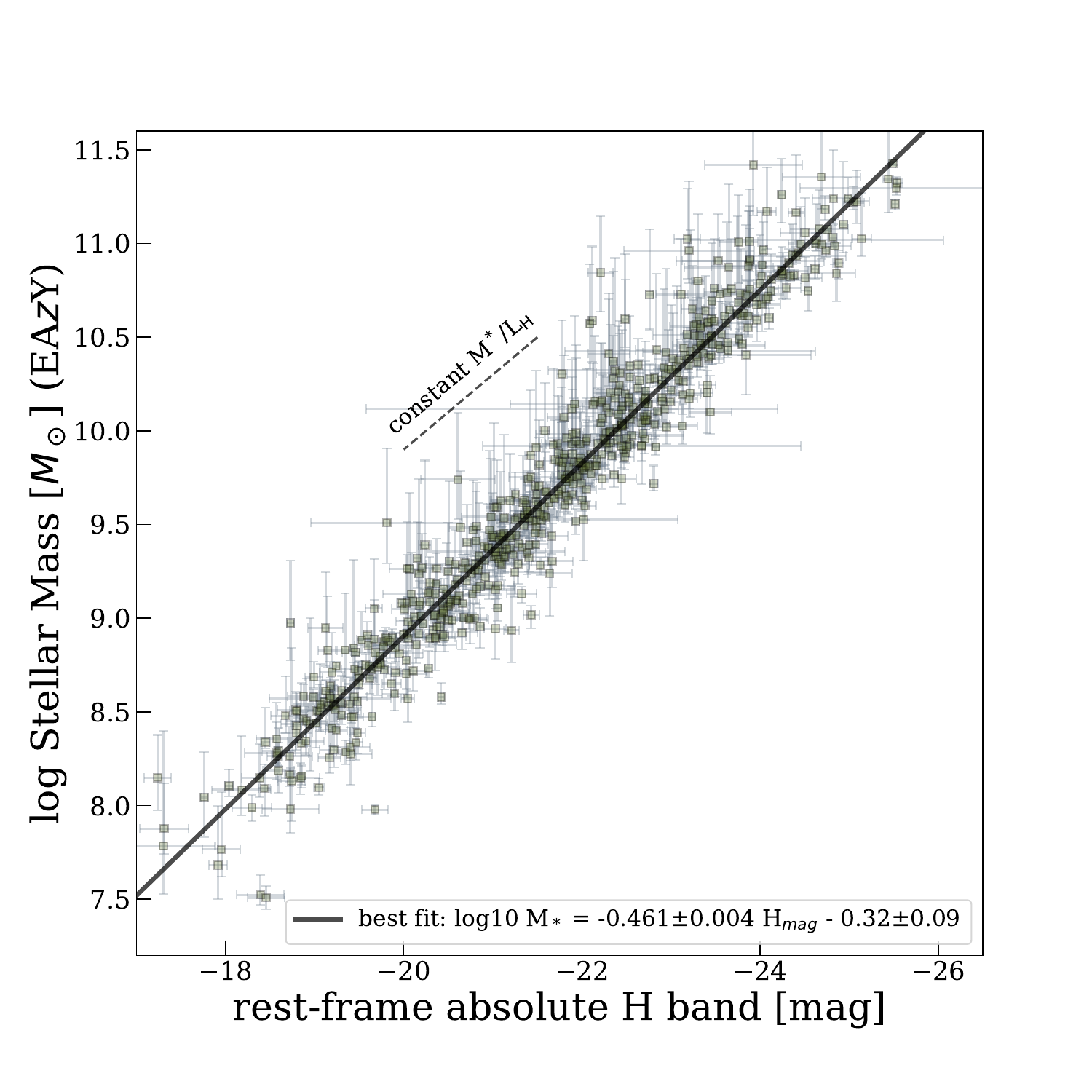}
    \caption{Rest-frame H band magnitude vs. log stellar mass as inferred from EA$z$Y SED fitting using photometric data across a wide range of wavelengths ($0.4-5.0\mu$m), along with 1$\sigma$ errorbars on both variables for the best fit model. The best fit linear relation (using bootstrapping and 3$\sigma$ outlier removal) is shown in black, with a slope of $\sim$0.46, implying a rest-frame H-band mass-to-light ratio that is not constant with mass.}
    \label{fig:fig3}
\end{figure}

We show the tight correlation between stellar mass and rest-frame H band magnitude in Figure \ref{fig:fig3}, as well as the best fit linear relation between the two properties. We note that the stellar masses are derived from SED fitting using photometric data across a wide range of wavelengths ($0.4-5.0\mu$m) and thus are not determined solely by the rest-frame H band.  {We perform linear regression with bootstrapping and 3$\sigma$ outlier removal and obtain the following equation that directly gives stellar mass from rest-frame H band magnitude:}

\begin{equation}
    \log(M^*)[M_\odot]=(-0.461\pm0.004)H_{\text{mag}}-(0.32\pm0.09),
\end{equation}

where ${H}_{\mathrm{mag}}$ is the absolute magnitude at rest-frame 1.6 micron. There is little scatter in the rest-frame H-band M/L ratio at fixed mass, with a biweight scatter around the best-fit of 0.17 dex,  {characterized by the biweight scale \citep{biweight1990} of the residuals around the best fit line.} A slope of -0.4 would imply a constant rest-frame H band mass-to-light ratio (M/L), but we find a slightly steeper slope, indicating that the M/L in the H-band is a function of mass and that not all of the galaxies in our Paschen-$\alpha$ sample have identical mass-to-light ratios in the rest-frame H band.

\subsection{Paschen-$\alpha$ Star Formation Rates}\label{sec:sfrpaa}
We use a relatively simple model for calculating SFR (see also, e.g., \citealp{papovich2009} and \citealp{cleri2022} for similar Paschen line conversions) and discuss further complexities that may be introduced in the SFR conversion in Section \ref{sec:uncertainties}. We derive P$\alpha$ SFRs using the procedure outlined in \cite{kennicutt2012} for H$\alpha$, adopting the atomic ratio between H$\alpha$ and Pa$\alpha$ to be 8.575 assuming Case B recombination and T=$10^4$K \citep{osterbrock1998}. We assume a Chabrier IMF and use the SFR$_{\mathrm{H}\alpha}$ calibration derived by \cite{murphy2011} and \cite{hao2011};  {we note that the choice between a \cite{kroupa2001} IMF used by these conversions and a \cite{chabrier2003} IMF has a negligible effect on the SFR calibration \citep{chomiuk2011,speagle2014}}. With our Pa $\alpha$ emission line measurements, we use the following conversion of L(Pa$\alpha$) to SFR$_{\text{Pa}\alpha}$:

\begin{equation}\label{eq:sfr}
    \text{SFR}_{\text{Pa}\alpha,\mathrm{corr}}[M_\odot/\text{yr}]=4.6\times10^{-41}\times L_{\text{Pa}\alpha,\mathrm{corr}} [\text{erg s$^{-1}$}]
\end{equation}

where $L_{\text{Pa}\alpha,\mathrm{corr}}$ is the dust-corrected Pa$\alpha$ luminosity, which we derive by applying a dust extinction correction to the measured emission line fluxes from FRESCO spectroscopic data and the luminosity distance using the grism redshifts for the default cosmology. The extinction in the V-band, A(V), is derived using EA$z$Y SED fitting, which we convert to an extinction correction for Pa$\alpha$ using a  {Cardelli attenuation law \citep{cardelli1989}.} We also calculate the corresponding extinction specifically around HII regions, A$_{\mathrm{V,HII}}$, found by \cite{price2014} to be A$_{\mathrm{V,HII}}$=1.86$^{+0.40}_{-0.37}$A$_{\mathrm{V,star}}$. We find that A(Pa$\alpha$) increases with galaxy mass, with a majority of A(Pa$\alpha$) values for sources in our sample falling below 0.2 mag, which corresponds to a relatively small extinction correction of $10^{0.4A(\text{Pa}\alpha)}\approx1.2$ that is applied to L(Pa$\alpha$). In comparison, using the same conversions for H$\alpha$ emission, A(H$\alpha$) would be about fives time greater than A(Pa$\alpha$), leading to significantly larger extinction corrections than those needed for Paschen-$\alpha$ emission. 

\begin{figure}
    \centering
    \includegraphics[width=\columnwidth]{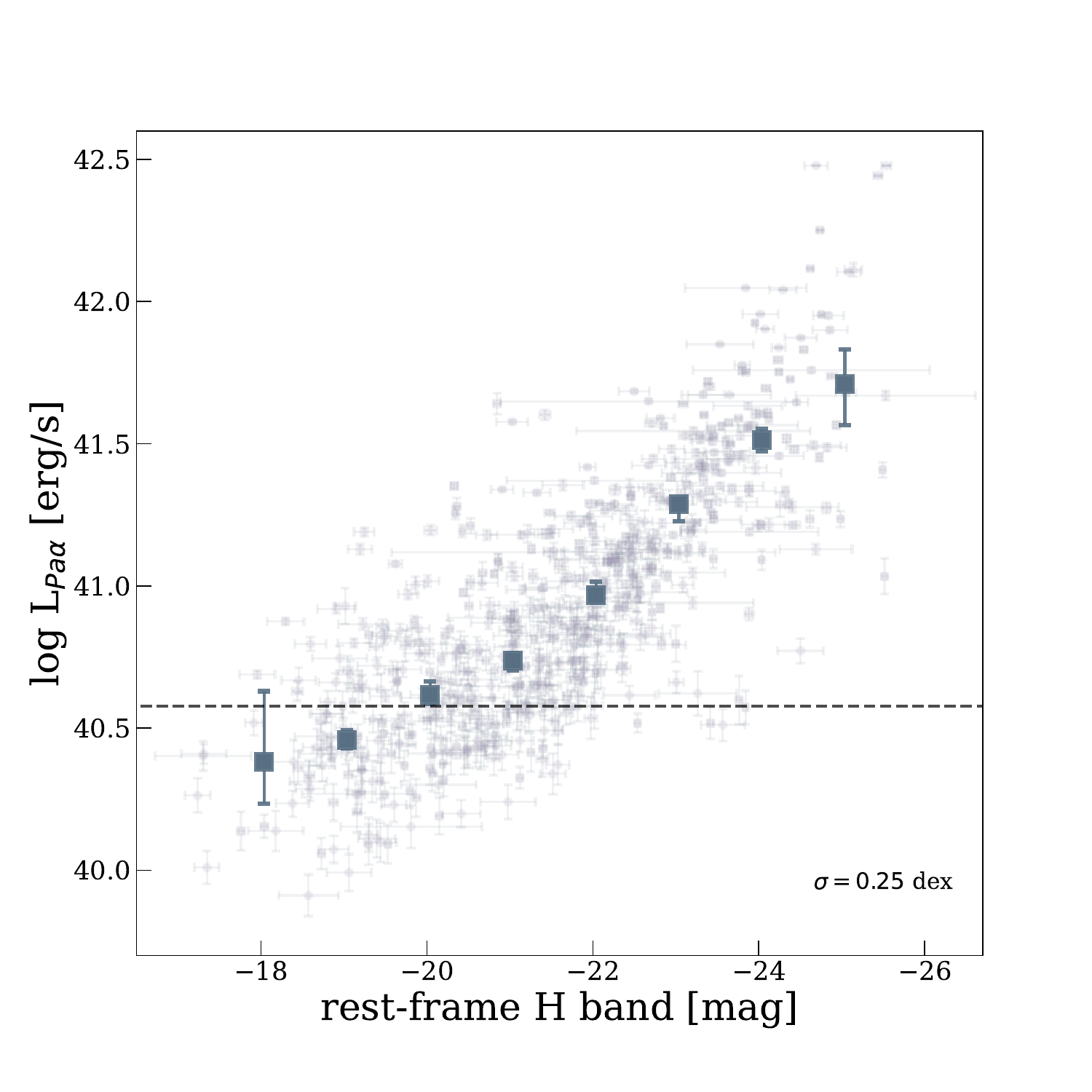}
    \caption{The directly observed relation between rest-frame H band absolute magnitude and Paschen-$\alpha$ luminosity, uncorrected for dust effects. Square points are median luminosities in H magnitude bins of width $\sim$1 mag, with with uncertainties estimated via bootstrap sampling in each bin. The biweight scatter of the relation is small at 0.25 dex. The horizontal dashed line corresponds to the Pa$\alpha$ flux detection limits for compact galaxies in our sample, below which we are not complete in luminosity, nor are we complete for larger galaxies.}
    \label{fig:fig4}
\end{figure}

The empirical relation between rest-frame H band magnitude (tracing stellar mass) and uncorrected Paschen-$\alpha$ luminosity (tracing SFR) is shown in Figure \ref{fig:fig4}.  {Binned median luminosities and uncertainties on these values are shown, obtained via bootstrap resampling in bins of 1 mag width,} and the biweight scatter in the relation is $\sim$0.25 dex. The completeness limits shown in this figure correspond to the detection limits for compact sources described in Section \ref{sec:data}. 

\subsection{The Star Forming Sequence}\label{sec:sfs}
\begin{figure*}
    \centering
    \includegraphics[width=\textwidth]{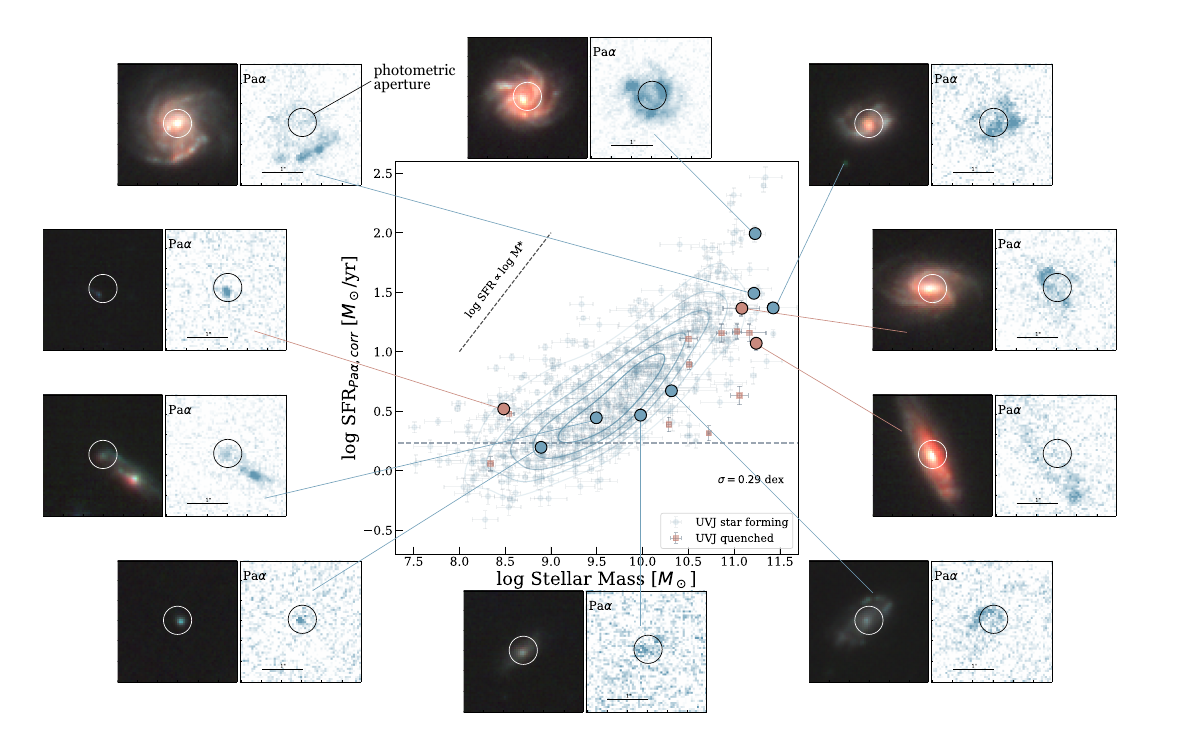}
    \caption{SFR, derived from Pa$\alpha$ emission line measurements, plotted against stellar mass, derived from EA$z$Y SED fitting. Galaxies in our sample are shown as either star forming (blue points) or quenched (red points), defined using the UVJ diagram in Figure \ref{fig:fig2}. The SFS scatter is similar to that found in the empirical relation between L(PaA) and rest-frame NIR magnitude. The dashed grey horizontal line indicates our completeness limit for compact sources, below which we cannot make any definitive statements on the SFR distributions. Spatially resolved Paschen-$\alpha$ emission line maps are shown for a selection of galaxies in our sample, with their locations relative to the distribution of data. For each galaxy, the left panel shows the RGB image, constructed by combining the F182M, F210M, and F444W bands, and the right panel shows the Paschen-$\alpha$ emission line map with 1" scaling shown. The size of the 0.7" circular aperture used for the photometric measurements are shown in both panels as well.}
    \label{fig:fig5}
\end{figure*}

We show the dust-corrected SFR (obtained from Pa$\alpha$ emission in Section \ref{sec:sfrpaa}) against the stellar mass (derived from EA$z$Y SED fitting that incorporates the rest-frame NIR bands as described in Section \ref{sec:massnir}), for our sample of galaxies in Figure \ref{fig:fig5}, where the horizontal dashed line indicates the SFR completeness limit. We calculate this from the flux detection limit for compact sources of $2\times10^{-18}$ erg s$^{-1}$, corresponding to a Paschen-$\alpha$ luminosity limit of $\sim10^{40.57}$ erg s$^{-1}$ at the highest redshift probed in our sample ($z=1.68$). We then convert this luminosity limit to a SFR limit based on the conversion in Equation \ref{eq:sfr}, noting that this is applicable for compact sources and thus we are not complete below this line nor are we complete for more extended sources.

\begin{figure*}
    \centering
    \includegraphics[width=0.7\textwidth]{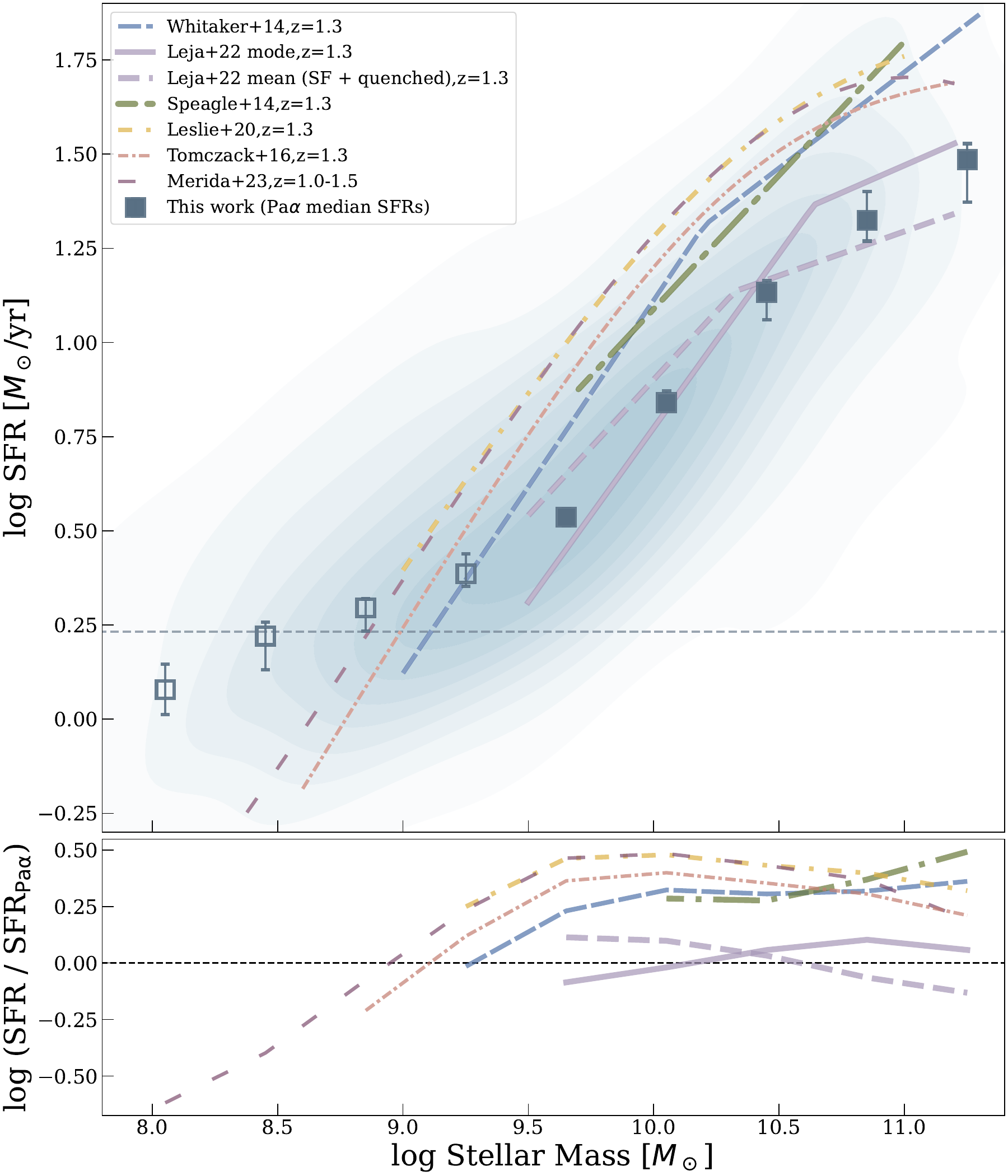}
    \caption{Top: Binned median points for our SFR-M* data compared to the relations found by \cite{speagle2014} (green), \cite{whitaker2014} (blue), \cite{tomczak2016} (red), \cite{leslie2020} (yellow), \cite{leja2022} (purple), \cite{merida2023} (magenta), with the horizontal dashed line indicating our SFR completeness limit and our distribution shown as a kernel density estimation. Bottom: Differences in SFRs between previous studies and our median values, with the horizontal dashed line indicating equality. We are offset from many earlier studies at intermediate masses and are most consistent with results from \cite{leja2022} at masses of $\sim10^{10-11}M_\odot$, where the solid and dashed purple lines indicate the mode (star forming sample only) and mean (star forming and quenched sample) relations from \cite{leja2022}.}
    \label{fig:fig6}
\end{figure*}

Quantifying a best fit line is nontrivial with scatter and uncertainties in both directions in the SFR-M* plane \citep{hogg2010}. While many studies fit the relation with a broken power law, with a flattened linear relation at higher masses (e.g., \citealp{whitaker2014,lee2015,leja2022}), we do not have enough data at the high mass end to motivate fitting a broken power law. We instead focus on analyzing the general trends we see here for the SFR-M* relation for galaxies in our sample. We calculate binned median values for mass bins of 0.4 dex width and present them in Figure \ref{fig:fig6}, with uncertainties in the SFR medians for each mass bin obtained from bootstrap resampling.

While we include all galaxies in our binned median SFRs, we color data points in  {Figure \ref{fig:fig5}} by whether a galaxy is considered star forming (blue) or quenched (red), as determined in Section \ref{sec:rfcolors} by position in UVJ space. It is interesting to note that several quenched galaxies have significant residual Paschen-$\alpha$ emission but were identified as quenched with the UVJ diagram. With slitless grism spectra, we have access to spatially resolved Paschen-$\alpha$ emission line maps, from which we can see where star formation,  {relatively} unobscured by dust, is occurring in these galaxies. In Figure \ref{fig:fig5}, we display several examples of these line maps in different regions throughout our SFR-M* distribution. One can see where star-forming clumps --- where Paschen-$\alpha$ emission is concentrated --- are present, and where our 0.7" circular photometric apertures are placed in relation. We return to this intriguing subsample of galaxies in Section \ref{sec:quenched}.

\subsection{Uncertainties and Caveats}\label{sec:uncertainties}
Here we explore caveats and uncertainties that may affect our measurements of the properties of the observed SFS. Our sample is most likely not complete at lower masses. We calculate the completeness limit of our sample, from detection and selection effects, and plot this as the dashed horizontal line in Figure \ref{fig:fig4} at $L_{\text{PaA}}=10^{40.57}$ erg s$^{-1}$. This corresponds to the 5$\sigma$ detection limit ($2\times10^{-18}$ erg s$^{-1}$) of the grism spectra at the highest redshift in our sample ($z=1.68$), below which we are most likely not complete. We note that this limit applies to compact galaxies, and the completeness is expected to be a function of size, with the sample likely being more complete for smaller galaxies and less complete for larger galaxies. This luminosity limit corresponds to SFR$_{\text{Pa}\alpha}=1.7$ M$_\odot$ yr$^{-1}$, which we show as the dashed horizontal line in Figure \ref{fig:fig5}. 

There are uncertainties associated with grism spectroscopy, such as overlapping spectra and spectral extraction  uncertainties, which may affect our measurements for SFR from Pa$\alpha$ emission lines; the  {measurement uncertainties associated with emission line fits} are incorporated into the errorbars shown for individual datapoints in Figure \ref{fig:fig5}. There are uncertainties associated with SED fitting, which may affect our measurements for M*; the 68\% confidence intervals are incorporated into the errorbars on the median stellar masses. 

We also note that, while Paschen-$\alpha$ is relatively dust-insensitive and the dust corrections that we measure are small, dust attenuation measurements obtained from SED fitting as well as the attenuation law used are a source of systematic uncertainty for our sample \citep{salim2020} and most likely depend on sSFR, stellar mass, and metallicity (e.g., \citealp{reddy2015,shivaei2020}). Additionally, the conversion between nebular line luminosity and SFR depends on the hardness of the ionizing spectrum, which in turn depends on the assumption of binary stars, stellar rotation, stellar metallicity, the chosen stellar population model, and the IMF. We assume a Solar metallicity calibration, which would be different for low metallicity and mass (see, e.g., \citealp{reddy2023} for a mass- and metallicity-dependent conversion,  {where the subsolar metallicity conversion factor is 0.37 dex lower than the factor used in this work}). Both of the main systematics (assuming binary stars and/or using lower metallcities) would lower our inferred SFRs further and we thus note that our SFR measurements are upper bounds. We emphasize that a main point of this paper is to directly observe the SFS at cosmic noon with the use of a novel dataset containing direct tracers of SFR (Paschen-$\alpha$ luminosity) and stellar mass (rest-frame H band magnitudes), and we use this direct conversion between Paschen-$\alpha$ luminosity and SFR.

Additionally, we do not fit the SFS as a function of redshift. Our redshift bin is thus larger than most used in previous studies ($z=$1.1-1.7), which can affect the shape of the relation (e.g., \citealp{popesso2023}). However, our sample has relatively few galaxies at $z>1.4$, and when comparing to previous studies we use their result at a redshift or range of redshifts similar to the majority of our sample ($z\sim1.3$ and $z=1.0-1.5$). Previous studies find that the normalization of the SFS evolves roughly as $(1+z)^{2.8}$ \citep{speagle2014,popesso2023}. Performing a differential redshift correction --- shifting the SFR of each source based on this redshift evolution in normalization --- reduces the biweight scatter in our relation by only $\sim$0.02 dex, and the binned median SFRs shift by at most 0.05 dex.

\section{Discussion}\label{sec:discussion}

\subsection{Scatter}\label{sec:scatter}
It is interesting to explore the different features in the structure of our SFR-M* data. In Figure \ref{fig:fig5}, one can see the general trend of how galaxies fall along and scatter around the SFS when using star formation rates and masses derived from our rest-frame NIR observations. We also show a comparison to a line of unity to demonstrate how different parts of the distribution deviate from a 1-1 relation between log(SFR) and log(M*), though we cannot make definitive statements about the slope of our relation at the low mass end due to completeness limits. 

The average binned biweight scatter is around 0.29 dex --- there is a tight relation between SFR-M*, indicating that most galaxies have similar star formation processes. Surprisingly, the scatter is 0.04 dex \textit{larger} than the purely empirical relation between Paschen-$\alpha$ luminosity and rest-frame H band magnitude found in Section \ref{sec:deriving}. The increase in scatter is likely largely due to the assumptions introduced in the conversions to SFR  {(i.e., dust corrections)} and stellar mass. This, in turn, suggests that the freedom in M/L ratios in the model fits to the data is larger than the variation in the actual universe.

\subsection{Comparison to Previous Studies}\label{sec:comparison}
Directly comparing the measured relations between SFR and M* is not trivial - \cite{speagle2014} find about 0.25-0.5 dex offset between relations in the literature due to various reasons (e.g., different SFR indicators and effects of preselection of star forming galaxies). Here, we attempt to compare our data with previous studies, specifically focusing on the use of this independent method of obtaining SFRs and stellar masses.

\subsubsection{Slope and Scatter}\label{sec:slopescattercomp}
In Figure \ref{fig:fig6}, we show our median datapoints and parametric relations for the SFS found by other studies; solid points indicate where our sample is complete. We compare our SFR-M* relation to six other studies, including \cite{whitaker2014} and \cite{leja2022}, which use 3D-HST data to measure SFRs using UV+IR and state-of-the-art SED fitting (\textsc{Prospector}), respectively. Additional studies shown in this figure use deep-IR observations to construct the SFR-M* relation (\citealp{tomczak2016}), rest-frame UV data \citep{merida2023}, or radio data \citep{leslie2020}. We also include the relation found by \cite{speagle2014}, which combines relations by several studies using mixed methods. 

We compare with relations that apply to the average redshift of our sample (Figure \ref{fig:fig1}), and we correct these relations to a Chabrier IMF to match this work. We do not include the offset factor of 0.3 dex for the stellar masses found by \cite{leja2022} as \cite{popesso2023} does in their comparisons, as we find masses consistent with those obtained using the \textsc{Prospector} framework (see Figure \ref{fig:fig7} and Appendix \ref{appendix:mass}).

We measure a scatter in SFRs at fixed mass of 0.29 dex, which is similar to that found by previous studies  {(e.g., \citealp{noeske2007,whitaker2012,speagle2014,leja2022})}. It is important to keep in mind that Paschen-$\alpha$ SFRs probe shorter timescales, similarly to H$\alpha$ emission line tracers \citep{kennicutt2012}, as compared to UV+IR SFR timescales. Paschen-$\alpha$ emission line tracers are also sensitive to the  {assumed IMF}, dust attenuation curve, and the assumption of the hardness of the ionizing radiation field when converting nebular line fluxes. The similar scatter with literature measurements then implies that there is no evidence for strong stochasticity on $\sim$10 Myr timescales that would be averaged out when using UV+IR tracers that are sensitive only to $\sim$100 Myr timescales.

Most studies predict a steeper relation at lower masses (e.g., \cite{merida2023} measure the SFS down to $10^8M_\odot$ at $z=1$, finding a slope of 0.97) and a flattening at high masses (e.g., with the flattening at the high mass end found by \cite{whitaker2014} and \cite{leja2022}). Comparing the slope, we can see that our relation has a shallower slope at lower masses and a steeper slope (closer to unity) at higher masses. However, we cannot completely quantify the slope below SFR$_{\text{Pa}\alpha}=1.7$ M$_\odot$ yr$^{-1}$ at the low mass end because of our detection-limited sample.  

On the bottom panel of Figure \ref{fig:fig6}, we show the differences between previous relations and our median SFR data points at our binned mass values. The black dashed line is where our data is equal to the compared data. At $M>9.5M_\odot$, we are offset from many earlier studies, with a relation that is up to 0.6 dex lower. Our results are the least offset with recent results obtained in the \textsc{Prospector} fitting framework \citep{leja2022}, with a mean difference of 0.09$\pm$0.04 dex at $M*\sim10^{10}-10^{11}M_\odot$ when comparing to the mean relation (dashed purple line in Figure \ref{fig:fig6}). 

The main point of this paper is not necessarily to more correctly characterize the SFS, but rather to probe a simpler, more independent method of determining the general shapes and trends in the relation without relying on the more complicated methods used in the studies shown here. There is not currently an observationally motivated way to tell whether or not this simpler method performs better or worse. We find that our relation is different than the relation predicted in other studies (lower measured SFRs or higher inferred masses), but we have agreement with \cite{leja2022} at the high mass end. As noted in Section \ref{sec:scatter}, the method that produces the lowest scatter would be to apply direct comparisons to Paschen-$\alpha$ luminosity and rest-frame H band magnitude.

\subsubsection{GOODS Fields SFRs and Masses}\label{sec:GOODSsfrscomp}

\begin{figure}
    \centering
    \includegraphics[width=\columnwidth]{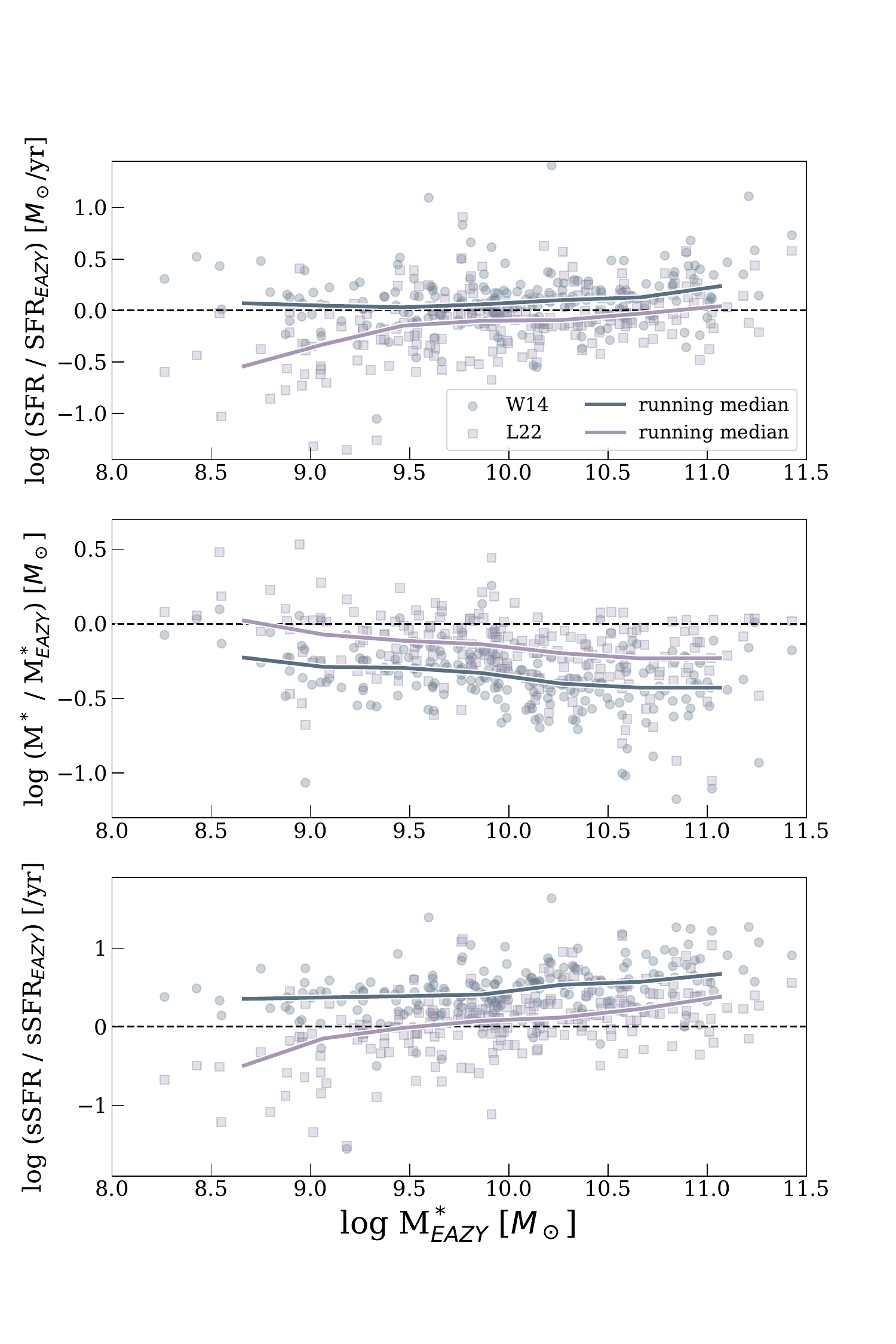}
    \caption{Comparison of SFRs and stellar masses for the same galaxies, cross-matched between FRESCO and 3D-HST catalogs with $\Delta z<0.05$, in the GOODS fields. We plot the difference between our measurements for SFR (top), stellar mass (middle), and sSFR (bottom) and the measurements from \textsc{Prospector} SED fitting \citep{leja2022} and FAST \citep{whitaker2014} in purple and blue, respectively. We plot the running medians for the distributions in the differences as solid lines, and dashed horizontal lines indicate equality. We are consistent in SFR with both previous studies, but our stellar masses from EA$z$Y are more consistent with those from \textsc{Prospector}, with an offset of 0.1-0.2 dex. Our stellar masses are further offset by $\sim0.3$ dex from those obtained with FAST.}
    \label{fig:fig7}
\end{figure}

We can isolate uncertainties associated with stellar mass and SFR measurements by looking directly at how different methods fare for the same sources in the GOODS fields. Various studies have constructed the SFS using the GOODS fields, including \cite{whitaker2014} and \cite{leja2022}, who measure SFRs from UV+IR and masses from FAST SED fitting  {\citep{kriek2009}}, and SFRs and masses from \textsc{Prospector} SED fitting, respectively. Because we measure the SFRs and masses for the same sample of galaxies in the GOODS-N and GOODS-S fields, we can directly compare our method of using Paschen-$\alpha$ emission as a SFR indicator and SED fitting incorporating rest-frame NIR bands as a stellar mass indicator. We show the differences between these measurements (\citealp{leja2022} in purple and \citealp{whitaker2014} in blue) for cross-matched sources with $\Delta z<0.05$ in Figure \ref{fig:fig7}, plotted against our measured masses. 

The top panel in Figure \ref{fig:fig7} shows comparisons in SFR measurements. We are consistent with both \cite{whitaker2014} and \cite{leja2022}, with mean differences of around 0.1 dex and a distribution that falls between the two previous studies. There is a larger deviation in the differences with \cite{leja2022} at the low mass end; previous studies have found an increase in scatter between different SFR indicators for galaxies at the lower end of the SFR distribution. \cite{cleri2022} find this when comparing between UV and Pa$\beta$ measurements at low redshift ($z\leq0.3$), and \cite{reddy2023} find at $z>1$ that this effect also exists when comparing UV SFRs to Paschen lines. 

We compare our EA$z$Y mass measurements, which incorporate rest-frame NIR photometry, to those obtained using SED fitting with UV-IR photometry from the 3D-HST survey \citep{brammer2012,skelton2014}, for the same sources in the GOODS fields in middle panel. We find that we are more consistent with the \textsc{Prospector} framework, with a difference of 0.0-0.2 dex across the entire mass range and a biweight scatter of $\sim$0.2 dex. We consistently measure higher masses than those measured using FAST SED fitting by $\sim$0.2-0.4 dex. This comparison between the masses of individual sources can explain why our SFR-M* relation has better agreement with \cite{leja2022} and is around 0.3 dex lower than the relation found by \cite{whitaker2014}. \cite{leja2019} find that the difference between stellar masses inferred from \textsc{Prospector} and FAST are due to the nonparametric SFHs in the \textsc{Prospector}-$\alpha$ template, which produces older stellar ages and favors a smooth SFH. FAST is sensitive to luminosity weighted ages and produces a large population of $\lesssim$ 1 Gyr old galaxies at this redshift (see Appendix \ref{appendix:mass} for further discussion). 

Our inferred stellar masses thus contribute to the offset in the SFS with many previous studies: when comparing sSFR measurements (bottom panel of Figure \ref{fig:fig7}, we are most consistent with the \textsc{Prospector} masses, and our offset in sSFR from \cite{whitaker2014} stems mostly from the offset in stellar mass. We find that our SFRs are consistent within the distributions of both \cite{whitaker2014} and \cite{leja2022} despite using a relatively simple conversion between Paschen-$\alpha$ and SFR.

\subsection{UVJ Quenched Galaxies with Paschen-$\alpha$ Emission}\label{sec:quenched}
We separate the star-forming and quenched galaxies in our sample to explore the distinguishing properties between the two regimes and the efficacy of the UVJ diagram. We can see where different populations of galaxies fall along the general shape of the SFS in Figure \ref{fig:fig5}, and there are interesting implications when looking at the locations of classically defined quenched galaxies along the SFS. While we do not have enough data points to fit a broken power law at the high mass end, we do see that most of the UVJ quenched galaxies, which tend to drive down the median SFR, fall within the high mass, low sSFR regime. This may be important to consider when fitting the relation to a selection of star-forming galaxies, especially if using the UVJ method: when not including the UVJ quenched galaxies in our star-forming selection, our median SFR values in the highest mass bins increase by up to 0.05 dex; while not significant for the case of this relatively small quenched sample, this may be important for larger surveys and cause an artificially high normalization when using UVJ selection criteria at these masses. This is especially seen in the comparisons between the mean and mode relations from \cite{leja2022} in Figure \ref{fig:fig6}.  {There are additionally a small number of UVJ quenched galaxies at the low mass, high sSFR end of the star forming sequence; the spatially resolved line map of one such example is shown in Figure \ref{fig:fig5}. While the circular aperture does fully incorporate the Paschen-$\alpha$ emission from the galaxy, it is still identified as quenched. This object likely falls in the category of being a dusty star forming galaxy that is misclassified by UVJ selection methods, which is an increasingly prevalent source of contamination at low masses (e.g., \citealp{diazgarcia2019}).}

As seen in  {Figure \ref{fig:fig5}}, several quenched galaxies also fall near the binned median SFR values, indicating that these galaxies have a significant amount of Paschen-$\alpha$ emission. Thus the bimodal classification system may not be reliable with the use of small, centrally-focused apertures that miss star formation in the outskirts. Looking at the spatially resolved line maps in Figure \ref{fig:fig5} of a subsample of galaxies, we can see that in most of the star-forming galaxies, star formation is concentrated in the central regions, whereas for some of the quenched galaxies, star formation is occurring mostly on the outskirts and is missed with the 0.7" circular apertures used in the photometric measurements. This could explain why some galaxies with significant Paschen-$\alpha$ emission are defined as quenched: their central regions may be dusty or are no longer star-forming, and thus in a small enough circular aperture, their color gradient would indicate quiescence. With an aperture size of 0.7" diameter, we detect 13 UVJ quenched galaxies; when decreasing the aperture size to a diameter of 0.32", we record an additional 5 UVJ quenched galaxies while gaining no star forming galaxies. Further explorations of color gradients and photometric aperture sizes in relation to the bimodal classification system are needed to improve global definitions of quenched galaxies. 

\section{Conclusion}\label{sec:conclusion}
With the capabilities of \textit{JWST}, we can now use Paschen-$\alpha$ as an independent,  {relatively} dust-insensitive tracer of SFR at cosmic noon in addition to incorporating rest-frame NIR photometry in derivations of stellar mass. This allows for a direct, independent measurement of the star forming sequence of galaxies at $z\sim1-2$. We summarize our main results as follows:

\begin{itemize}
    \item We find that there is low scatter (0.25 dex) in the directly observed relation between the rest-frame H band magnitude --- which traces the dominant stellar mass component in a galaxy --- and dust-uncorrected Paschen-$\alpha$ luminosity --- which traces SFR. The scatter increases slightly in the relation between log(SFR) and log(M*) (likely due to assumptions introduced in the conversions to stellar mass and  {dust-corrected} SFR) and is consistent with previous studies at $\sim0.29$ dex. Various assumptions used and disagreements in masses inferred from different SED codes lead to increased scatter among SFR-M* relations in the literature, and we note that the scatter in the relationship between Paschen-$\alpha$ luminosity and rest-frame H band magnitude is much smaller than these systematic differences.
    \item {At $\sim10^{10-11} M_\odot$, our relation is offset from most studies. It is in good agreement with results obtained using the \textsc{Prospector} fitting framework \citep{leja2022}; we find that this is mostly attributed to the similarities between the masses obtained from the EA$z$Y redshift-dependent SFH templates and the nonparametric SFHs implemented in \sc{Prospector}.}
    \item Some UVJ defined quenched galaxies fall close to the relation, and spatially resolved emission line maps obtained from grism spectroscopic data allow us to probe exactly where star formation is occurring in these galaxies. The bimodal system of selecting star forming galaxy is highly sensitive to color gradients, and it is thus important for larger surveys of the SFS to use carefully selected photometric aperture sizes. 
\end{itemize}

With more incoming data from \textit{JWST}, this method will be able to expand our understanding of the SFR-M* relation through simpler, independent measurements of SFR and stellar mass at cosmic noon. The process of quenching can be further explored by looking at the morphology of Paschen-$\alpha$ emission using emission line maps produced by grism spectroscopy. Our methods can also be applied to larger samples to better measure the SFS in this redshift regime and constrain the low mass end of the relation.

\section*{Acknowledgments}

 {The authors thank the anonymous referee whose comments and suggestions improved the quality of this work.}

This work is based on observations made with the NASA/ESA/CSA James Webb Space Telescope. The data were obtained from the Mikulski Archive for Space Telescopes at the Space Telescope Science Institute, which is operated by the Association of Universities for Research in Astronomy, Inc., under NASA contract NAS 5-03127 for JWST. These observations are associated with program \# 1895.

Support for this work was provided by NASA through grant JWST-GO-01895 awarded by the Space Telescope Science Institute, which is operated by the Association of Universities for Research in Astronomy, Inc., under NASA contract NAS 5-26555.

This work has received funding from the Swiss State Secretariat for Education, Research and Innovation (SERI) under contract number MB22.00072, as well as from the Swiss National Science Foundation (SNSF) through project grant 200020\_207349.
The Cosmic Dawn Center (DAWN) is funded by the Danish National Research Foundation under grant No.\ 140.

RPN acknowledges funding from JWST programs GO-1933 and GO-2279. Support for this work was provided by NASA through the NASA Hubble Fellowship grant HST-HF2-51515.001-A awarded by the Space Telescope Science Institute, which is operated by the Association of Universities for Research in Astronomy, Incorporated, under NASA contract NAS5-26555

Cloud-based data processing and file storage for this work is provided by the AWS Cloud Credits for Research program.

This paper made use of several publicly available software packages. We thank the respective authors for sharing their work: \texttt{IPython} \citep{ipython},
    \texttt{matplotlib} \citep{matplotlib},
    \texttt{seaborn} \citep[][]{seaborn},
    \texttt{numpy} \citep{numpy},
    \texttt{scipy} \citep{scipy},
    \texttt{jupyter} \citep{jupyter},
    \texttt{Astropy}
    \citep{astropy:2013,astropy:2018,astropy:2022},
    \texttt{grizli}
    (\citealt{grizli,grizli2}),
   \texttt{Prospector} \citep[][]{leja2019,leja2017,johnson2021},
   \texttt{FSPS} \citep[][]{FSPS1,FSPS2,FSPS3,FSPS4,python-FSPS},
   \texttt{dynesty} \citep[][]{dynesty},
    \texttt{EAZY} \citep[][]{eazy},
    \texttt{Bagpipes} \citep[][]{carnall2018},
    \texttt{SExtractor} \citep[][]{Bertin96}.
    
\section*{Data Availability}
The JWST data presented in this article were obtained from the Mikulski Archive for Space Telescopes (MAST) at the Space Telescope Science Institute. The specific observations analyzed can be accessed via \dataset[DOI: 10.17909/gdyc-7g80]{https://doi.org/10.17909/gdyc-7g80}.

\appendix

\section{Mass measurements}\label{appendix:mass}

In this work, we infer masses using EA$z$Y SED templates due to the ease of use of the code and the reproducibility of our results, as discussed in Section \ref{sec:massnir}. While several previous studies use similar EA$z$Y templates to infer masses (see, e.g., \citealp{sherman2020,labbe2023}), we acknowledge that EA$z$Y has not been traditionally used in the context of inferring physical properties of galaxies. We thus compare the stellar masses used in this work to stellar masses from other SED fitting codes, namely, \textsc{Bagpipes} \citep{carnall2018} and \textsc{Prospector} \citep{johnson2021}. For all fits, we use grism redshifts and 0.7" photometric apertures. 

We specifically use {Prospector}-$\alpha$, which includes non-paramtetric SFHs, nebular emission treatment, and stellar population synthesis (SPS) models using \textsc{FSPS} (\citealp{conroy2009,conroy2010}), following the parameters outlined in \cite{leja2019}. We use a delayed tau model with SPS models from \cite{bruzual2003} for the {Bagpipes} runs. For the Prospector masses, we check the chi squared value for each fit and discard those with bad fits from our results.

\begin{figure}
    \centering
    \includegraphics[width=\columnwidth]{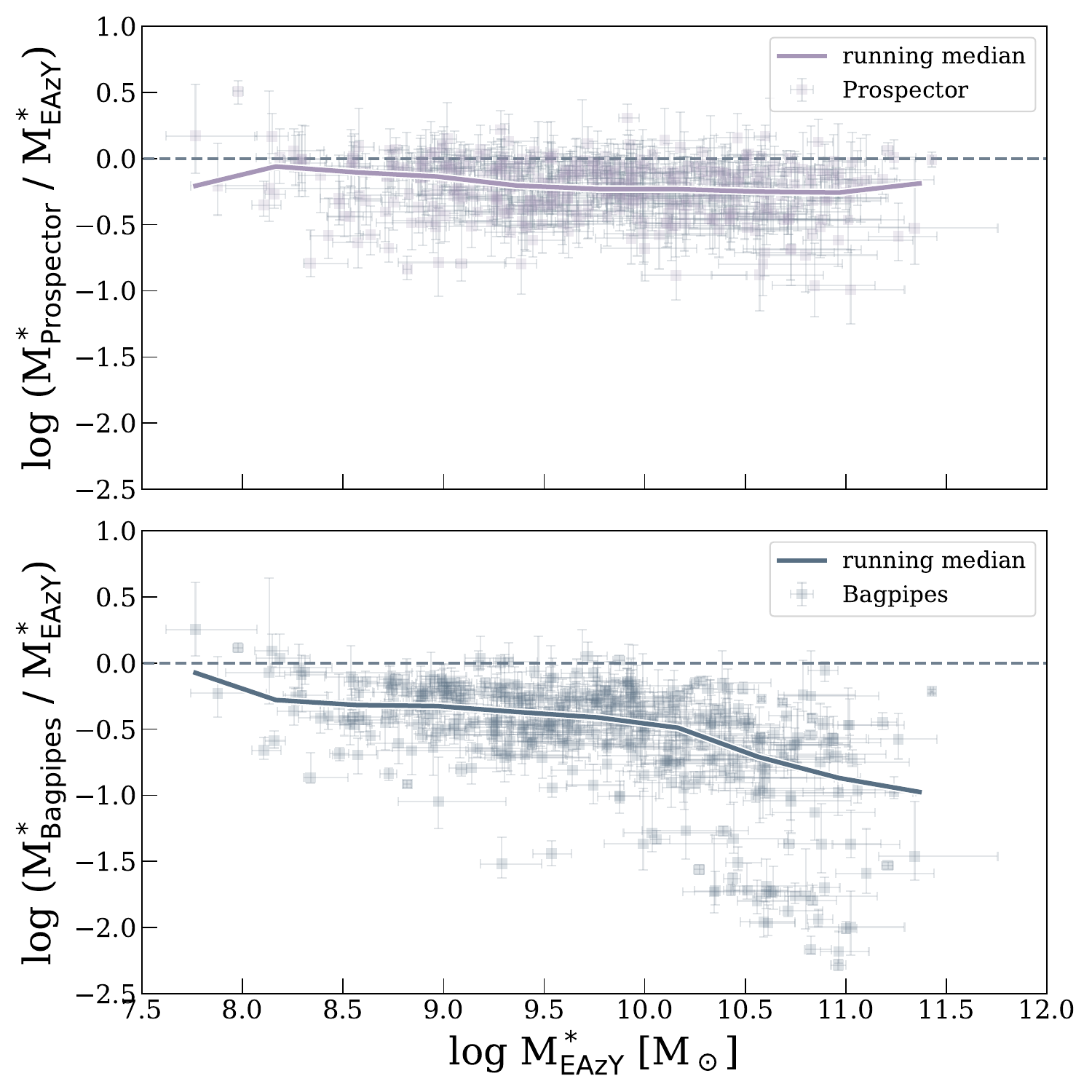}
    \caption{Stellar mass comparisons between the masses used in this work, derived with EA$z$Y, and masses derived using Prospector (top) and Bagpipes (bottom). We are more consistent with masses derived with Prospector (0.2 dex median offset) because both template superposition (EA$z$Y) and non-parametric SFHs (Prospector-$\alpha$) allow for multi-component SFHs, whereas SED fitting codes such as Bagpipes and FAST implement parametric SFHs.}
    \label{fig:fig8}
\end{figure}

\begin{figure}
    \centering
    \includegraphics[width=\columnwidth]{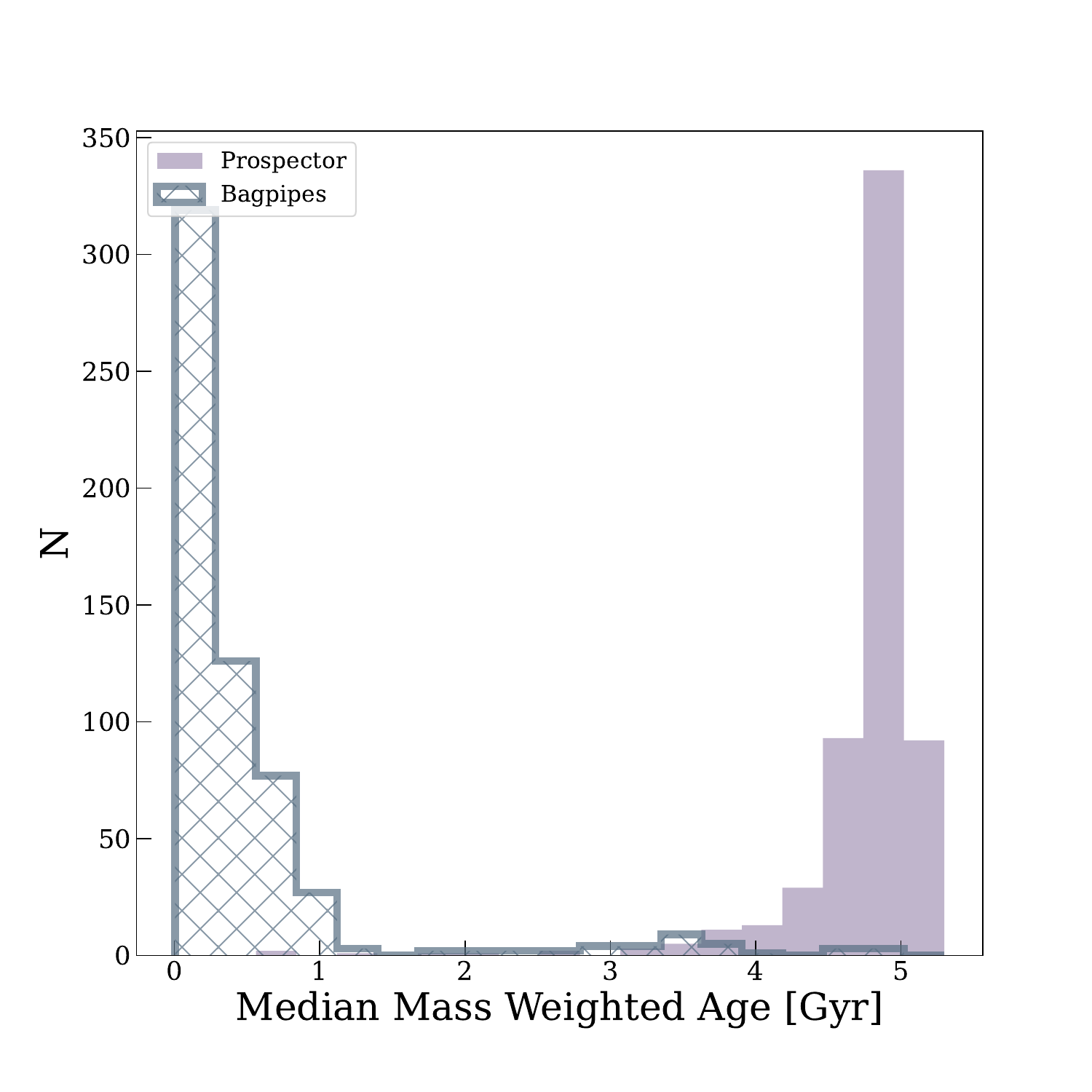}
    \caption{Comparisons between the mass-weighted ages derived using Prospector (solid purple) and Bagpipes (hatched blue). Prospector includes the mass contribution from an older population of stars, thus resulting in older mass-weighted ages.}
    \label{fig:fig9}
\end{figure}

Our comparison is shown in Figure \ref{fig:fig8}, with differences between the EA$z$Y stellar masses used in this work and stellar masses from Prospector (top) and Bagpipes (bottom). Our masses are higher than the results from each of the other SED fitting codes, but we are most consistent with Prospector with a median offset of 0.20 dex and a biweight scatter of 0.19 dex, which is around the scatter expected between different SED fitting codes (see, e.g., \citealp{pacifici2023} for a discussion on the scatter in physical properties between different SED fitting codes), whereas we measure higher masses than Bagpipes by a median offset of 0.46 dex and 0.35 biweight scatter. These results are indicative of the differences in the SFH implementations in these codes; the Prospector-$\alpha$ model implements non-paramtric SFHs and captures the mass contribution from older stellar populations (e.g., \citealp{leja2019,lower2020}), while Bagpipes, like FAST, is sensitive to the youngest stars. This is especially evident in Figure \ref{fig:fig9}: we see that the mass-weighted ages from Bagpipes are significantly and consistently lower than those derived from Prospector, implying that Bagpipes models a much younger stellar population than Prospector. Using SED template superposition with EA$z$Y, which allows for multi-component SFHs rather than a unimodel SFH, we also infer higher stellar masses and thus obtain similar results to Prospector. 

\bibliography{paa_sfs}{}

\begin{thebibliography}{}
\expandafter\ifx\csname natexlab\endcsname\relax\def\natexlab#1{#1}\fi
\providecommand{\url}[1]{\href{#1}{#1}}
\providecommand{\dodoi}[1]{doi:~\href{http://doi.org/#1}{\nolinkurl{#1}}}
\providecommand{\doeprint}[1]{\href{http://ascl.net/#1}{\nolinkurl{http://ascl.net/#1}}}
\providecommand{\doarXiv}[1]{\href{https://arxiv.org/abs/#1}{\nolinkurl{https://arxiv.org/abs/#1}}}

\bibitem[{{Abramson} {et~al.}(2014){Abramson}, {Kelson}, {Dressler}, {Poggianti}, {Gladders}, {Oemler}, \& {Vulcani}}]{abramson2014}
{Abramson}, L.~E., {Kelson}, D.~D., {Dressler}, A., {et~al.} 2014, \apjl, 785, L36, \dodoi{10.1088/2041-8205/785/2/L36}

\bibitem[{{Astropy Collaboration} {et~al.}(2013){Astropy Collaboration}, {Robitaille}, {Tollerud}, {Greenfield}, {Droettboom}, {Bray}, {Aldcroft}, {Davis}, {Ginsburg}, {Price-Whelan}, {Kerzendorf}, {Conley}, {Crighton}, {Barbary}, {Muna}, {Ferguson}, {Grollier}, {Parikh}, {Nair}, {Unther}, {Deil}, {Woillez}, {Conseil}, {Kramer}, {Turner}, {Singer}, {Fox}, {Weaver}, {Zabalza}, {Edwards}, {Azalee Bostroem}, {Burke}, {Casey}, {Crawford}, {Dencheva}, {Ely}, {Jenness}, {Labrie}, {Lim}, {Pierfederici}, {Pontzen}, {Ptak}, {Refsdal}, {Servillat}, \& {Streicher}}]{astropy:2013}
{Astropy Collaboration}, {Robitaille}, T.~P., {Tollerud}, E.~J., {et~al.} 2013, \aap, 558, A33, \dodoi{10.1051/0004-6361/201322068}

\bibitem[{{Astropy Collaboration} {et~al.}(2018){Astropy Collaboration}, {Price-Whelan}, {Sip{\H{o}}cz}, {G{\"u}nther}, {Lim}, {Crawford}, {Conseil}, {Shupe}, {Craig}, {Dencheva}, {Ginsburg}, {Vand erPlas}, {Bradley}, {P{\'e}rez-Su{\'a}rez}, {de Val-Borro}, {Aldcroft}, {Cruz}, {Robitaille}, {Tollerud}, {Ardelean}, {Babej}, {Bach}, {Bachetti}, {Bakanov}, {Bamford}, {Barentsen}, {Barmby}, {Baumbach}, {Berry}, {Biscani}, {Boquien}, {Bostroem}, {Bouma}, {Brammer}, {Bray}, {Breytenbach}, {Buddelmeijer}, {Burke}, {Calderone}, {Cano Rodr{\'\i}guez}, {Cara}, {Cardoso}, {Cheedella}, {Copin}, {Corrales}, {Crichton}, {D'Avella}, {Deil}, {Depagne}, {Dietrich}, {Donath}, {Droettboom}, {Earl}, {Erben}, {Fabbro}, {Ferreira}, {Finethy}, {Fox}, {Garrison}, {Gibbons}, {Goldstein}, {Gommers}, {Greco}, {Greenfield}, {Groener}, {Grollier}, {Hagen}, {Hirst}, {Homeier}, {Horton}, {Hosseinzadeh}, {Hu}, {Hunkeler}, {Ivezi{\'c}}, {Jain}, {Jenness}, {Kanarek}, {Kendrew}, {Kern}, {Kerzendorf}, {Khvalko}, {King}, {Kirkby}, {Kulkarni},
  {Kumar}, {Lee}, {Lenz}, {Littlefair}, {Ma}, {Macleod}, {Mastropietro}, {McCully}, {Montagnac}, {Morris}, {Mueller}, {Mumford}, {Muna}, {Murphy}, {Nelson}, {Nguyen}, {Ninan}, {N{\"o}the}, {Ogaz}, {Oh}, {Parejko}, {Parley}, {Pascual}, {Patil}, {Patil}, {Plunkett}, {Prochaska}, {Rastogi}, {Reddy Janga}, {Sabater}, {Sakurikar}, {Seifert}, {Sherbert}, {Sherwood-Taylor}, {Shih}, {Sick}, {Silbiger}, {Singanamalla}, {Singer}, {Sladen}, {Sooley}, {Sornarajah}, {Streicher}, {Teuben}, {Thomas}, {Tremblay}, {Turner}, {Terr{\'o}n}, {van Kerkwijk}, {de la Vega}, {Watkins}, {Weaver}, {Whitmore}, {Woillez}, {Zabalza}, \& {Astropy Contributors}}]{astropy:2018}
{Astropy Collaboration}, {Price-Whelan}, A.~M., {Sip{\H{o}}cz}, B.~M., {et~al.} 2018, \aj, 156, 123, \dodoi{10.3847/1538-3881/aabc4f}

\bibitem[{{Astropy Collaboration} {et~al.}(2022){Astropy Collaboration}, {Price-Whelan}, {Lim}, {Earl}, {Starkman}, {Bradley}, {Shupe}, {Patil}, {Corrales}, {Brasseur}, {N{"o}the}, {Donath}, {Tollerud}, {Morris}, {Ginsburg}, {Vaher}, {Weaver}, {Tocknell}, {Jamieson}, {van Kerkwijk}, {Robitaille}, {Merry}, {Bachetti}, {G{"u}nther}, {Aldcroft}, {Alvarado-Montes}, {Archibald}, {B{'o}di}, {Bapat}, {Barentsen}, {Baz{'a}n}, {Biswas}, {Boquien}, {Burke}, {Cara}, {Cara}, {Conroy}, {Conseil}, {Craig}, {Cross}, {Cruz}, {D'Eugenio}, {Dencheva}, {Devillepoix}, {Dietrich}, {Eigenbrot}, {Erben}, {Ferreira}, {Foreman-Mackey}, {Fox}, {Freij}, {Garg}, {Geda}, {Glattly}, {Gondhalekar}, {Gordon}, {Grant}, {Greenfield}, {Groener}, {Guest}, {Gurovich}, {Handberg}, {Hart}, {Hatfield-Dodds}, {Homeier}, {Hosseinzadeh}, {Jenness}, {Jones}, {Joseph}, {Kalmbach}, {Karamehmetoglu}, {Ka{l}uszy{'n}ski}, {Kelley}, {Kern}, {Kerzendorf}, {Koch}, {Kulumani}, {Lee}, {Ly}, {Ma}, {MacBride}, {Maljaars}, {Muna}, {Murphy}, {Norman}, {O'Steen},
  {Oman}, {Pacifici}, {Pascual}, {Pascual-Granado}, {Patil}, {Perren}, {Pickering}, {Rastogi}, {Roulston}, {Ryan}, {Rykoff}, {Sabater}, {Sakurikar}, {Salgado}, {Sanghi}, {Saunders}, {Savchenko}, {Schwardt}, {Seifert-Eckert}, {Shih}, {Jain}, {Shukla}, {Sick}, {Simpson}, {Singanamalla}, {Singer}, {Singhal}, {Sinha}, {Sip{H{o}}cz}, {Spitler}, {Stansby}, {Streicher}, {{{S}}umak}, {Swinbank}, {Taranu}, {Tewary}, {Tremblay}, {Val-Borro}, {Van Kooten}, {Vasovi{'c}}, {Verma}, {de Miranda Cardoso}, {Williams}, {Wilson}, {Winkel}, {Wood-Vasey}, {Xue}, {Yoachim}, {Zhang}, {Zonca}, \& {Astropy Project Contributors}}]{astropy:2022}
{Astropy Collaboration}, {Price-Whelan}, A.~M., {Lim}, P.~L., {et~al.} 2022, \apj, 935, 167, \dodoi{10.3847/1538-4357/ac7c74}

\bibitem[{{Atek} {et~al.}(2022){Atek}, {Furtak}, {Oesch}, {van Dokkum}, {Reddy}, {Contini}, {Illingworth}, \& {Wilkins}}]{atek2022}
{Atek}, H., {Furtak}, L.~J., {Oesch}, P., {et~al.} 2022, \mnras, 511, 4464, \dodoi{10.1093/mnras/stac360}

\bibitem[{{Beers} {et~al.}(1990){Beers}, {Flynn}, \& {Gebhardt}}]{biweight1990}
{Beers}, T.~C., {Flynn}, K., \& {Gebhardt}, K. 1990, \aj, 100, 32, \dodoi{10.1086/115487}

\bibitem[{{Belli} {et~al.}(2015){Belli}, {Newman}, \& {Ellis}}]{belli2015}
{Belli}, S., {Newman}, A.~B., \& {Ellis}, R.~S. 2015, \apj, 799, 206, \dodoi{10.1088/0004-637X/799/2/206}

\bibitem[{{Bertin} \& {Arnouts}(1996{\natexlab{a}})}]{sextractor}
{Bertin}, E., \& {Arnouts}, S. 1996{\natexlab{a}}, \aaps, 117, 393, \dodoi{10.1051/aas:1996164}

\bibitem[{{Bertin} \& {Arnouts}(1996{\natexlab{b}})}]{Bertin96}
---. 1996{\natexlab{b}}, \aaps, 117, 393

\bibitem[{{Boogaard} {et~al.}(2018){Boogaard}, {Brinchmann}, {Bouch{\'e}}, {Paalvast}, {Bacon}, {Bouwens}, {Contini}, {Gunawardhana}, {Inami}, {Marino}, {Maseda}, {Mitchell}, {Nanayakkara}, {Richard}, {Schaye}, {Schreiber}, {Tacchella}, {Wisotzki}, \& {Zabl}}]{boogaard2018}
{Boogaard}, L.~A., {Brinchmann}, J., {Bouch{\'e}}, N., {et~al.} 2018, \aap, 619, A27, \dodoi{10.1051/0004-6361/201833136}

\bibitem[{{Brammer}(2018)}]{grizli}
{Brammer}, G. 2018, {Gbrammer/Grizli: Preliminary Release}, 0.4.0, Zenodo,  Zenodo, \dodoi{10.5281/zenodo.1146905}

\bibitem[{{Brammer} {et~al.}(2022){Brammer}, {Strait}, {Matharu}, \& {Momcheva}}]{grizli2}
{Brammer}, G., {Strait}, V., {Matharu}, J., \& {Momcheva}, I. 2022, {grizli}, 1.5.0, Zenodo,  Zenodo, \dodoi{10.5281/zenodo.6672538}

\bibitem[{{Brammer} {et~al.}(2008){Brammer}, {van Dokkum}, \& {Coppi}}]{eazy}
{Brammer}, G.~B., {van Dokkum}, P.~G., \& {Coppi}, P. 2008, \apj, 686, 1503, \dodoi{10.1086/591786}

\bibitem[{{Brammer} {et~al.}(2011){Brammer}, {Whitaker}, {van Dokkum}, {Marchesini}, {Franx}, {Kriek}, {Labb{\'e}}, {Lee}, {Muzzin}, {Quadri}, {Rudnick}, \& {Williams}}]{brammer2011}
{Brammer}, G.~B., {Whitaker}, K.~E., {van Dokkum}, P.~G., {et~al.} 2011, \apj, 739, 24, \dodoi{10.1088/0004-637X/739/1/24}

\bibitem[{{Brammer} {et~al.}(2012){Brammer}, {van Dokkum}, {Franx}, {Fumagalli}, {Patel}, {Rix}, {Skelton}, {Kriek}, {Nelson}, {Schmidt}, {Bezanson}, {da Cunha}, {Erb}, {Fan}, {F{\"o}rster Schreiber}, {Illingworth}, {Labb{\'e}}, {Leja}, {Lundgren}, {Magee}, {Marchesini}, {McCarthy}, {Momcheva}, {Muzzin}, {Quadri}, {Steidel}, {Tal}, {Wake}, {Whitaker}, \& {Williams}}]{brammer2012}
{Brammer}, G.~B., {van Dokkum}, P.~G., {Franx}, M., {et~al.} 2012, \apjs, 200, 13, \dodoi{10.1088/0067-0049/200/2/13}

\bibitem[{{Bruzual} \& {Charlot}(2003)}]{bruzual2003}
{Bruzual}, G., \& {Charlot}, S. 2003, \mnras, 344, 1000, \dodoi{10.1046/j.1365-8711.2003.06897.x}

\bibitem[{{Bundy} {et~al.}(2010){Bundy}, {Scarlata}, {Carollo}, {Ellis}, {Drory}, {Hopkins}, {Salvato}, {Leauthaud}, {Koekemoer}, {Murray}, {Ilbert}, {Oesch}, {Ma}, {Capak}, {Pozzetti}, \& {Scoville}}]{bundy2010}
{Bundy}, K., {Scarlata}, C., {Carollo}, C.~M., {et~al.} 2010, \apj, 719, 1969, \dodoi{10.1088/0004-637X/719/2/1969}

\bibitem[{{Cardamone} {et~al.}(2010){Cardamone}, {Urry}, {Schawinski}, {Treister}, {Brammer}, \& {Gawiser}}]{cardamone2010}
{Cardamone}, C.~N., {Urry}, C.~M., {Schawinski}, K., {et~al.} 2010, \apjl, 721, L38, \dodoi{10.1088/2041-8205/721/1/L38}

\bibitem[{{Cardelli} {et~al.}(1989){Cardelli}, {Clayton}, \& {Mathis}}]{cardelli1989}
{Cardelli}, J.~A., {Clayton}, G.~C., \& {Mathis}, J.~S. 1989, \apj, 345, 245, \dodoi{10.1086/167900}

\bibitem[{{Carnall} {et~al.}(2018){Carnall}, {McLure}, {Dunlop}, \& {Dav{\'e}}}]{carnall2018}
{Carnall}, A.~C., {McLure}, R.~J., {Dunlop}, J.~S., \& {Dav{\'e}}, R. 2018, \mnras, 480, 4379, \dodoi{10.1093/mnras/sty2169}

\bibitem[{{Chabrier}(2003)}]{chabrier2003}
{Chabrier}, G. 2003, \pasp, 115, 763, \dodoi{10.1086/376392}

\bibitem[{{Chomiuk} \& {Povich}(2011)}]{chomiuk2011}
{Chomiuk}, L., \& {Povich}, M.~S. 2011, \aj, 142, 197, \dodoi{10.1088/0004-6256/142/6/197}

\bibitem[{{Cleri} {et~al.}(2022){Cleri}, {Trump}, {Backhaus}, {Momcheva}, {Papovich}, {Simons}, {Weiner}, {Estrada-Carpenter}, {Finkelstein}, {Giavalisco}, {Ji}, {Jung}, {Matharu}, {Martinez}, \& {Sturm}}]{cleri2022}
{Cleri}, N.~J., {Trump}, J.~R., {Backhaus}, B.~E., {et~al.} 2022, \apj, 929, 3, \dodoi{10.3847/1538-4357/ac5a4c}

\bibitem[{{Conroy} \& {Gunn}(2010{\natexlab{a}})}]{FSPS3}
{Conroy}, C., \& {Gunn}, J.~E. 2010{\natexlab{a}}, \apj, 712, 833, \dodoi{10.1088/0004-637X/712/2/833}

\bibitem[{{Conroy} \& {Gunn}(2010{\natexlab{b}})}]{FSPS4}
---. 2010{\natexlab{b}}, {FSPS: Flexible Stellar Population Synthesis}, Astrophysics Source Code Library, record ascl:1010.043.
\newblock \doeprint{1010.043}

\bibitem[{{Conroy} \& {Gunn}(2010{\natexlab{c}})}]{conroy2010}
---. 2010{\natexlab{c}}, \apj, 712, 833, \dodoi{10.1088/0004-637X/712/2/833}

\bibitem[{{Conroy} {et~al.}(2009{\natexlab{a}}){Conroy}, {Gunn}, \& {White}}]{FSPS1}
{Conroy}, C., {Gunn}, J.~E., \& {White}, M. 2009{\natexlab{a}}, \apj, 699, 486, \dodoi{10.1088/0004-637X/699/1/486}

\bibitem[{{Conroy} {et~al.}(2009{\natexlab{b}}){Conroy}, {Gunn}, \& {White}}]{conroy2009}
---. 2009{\natexlab{b}}, \apj, 699, 486, \dodoi{10.1088/0004-637X/699/1/486}

\bibitem[{{Conroy} {et~al.}(2010){Conroy}, {White}, \& {Gunn}}]{FSPS2}
{Conroy}, C., {White}, M., \& {Gunn}, J.~E. 2010, \apj, 708, 58, \dodoi{10.1088/0004-637X/708/1/58}

\bibitem[{{Cooke} {et~al.}(2023){Cooke}, {Kartaltepe}, {Rose}, {Tyler}, {Darvish}, {Leslie}, {Peng}, {H{\"a}u{\ss}ler}, \& {Koekemoer}}]{Cooke2023}
{Cooke}, K.~C., {Kartaltepe}, J.~S., {Rose}, C., {et~al.} 2023, \apj, 942, 49, \dodoi{10.3847/1538-4357/aca40f}

\bibitem[{{Curtis-Lake} {et~al.}(2021){Curtis-Lake}, {Chevallard}, {Charlot}, \& {Sandles}}]{curtislake2021}
{Curtis-Lake}, E., {Chevallard}, J., {Charlot}, S., \& {Sandles}, L. 2021, \mnras, 503, 4855, \dodoi{10.1093/mnras/stab698}

\bibitem[{{Daddi} {et~al.}(2007){Daddi}, {Dickinson}, {Morrison}, {Chary}, {Cimatti}, {Elbaz}, {Frayer}, {Renzini}, {Pope}, {Alexander}, {Bauer}, {Giavalisco}, {Huynh}, {Kurk}, \& {Mignoli}}]{daddi2007}
{Daddi}, E., {Dickinson}, M., {Morrison}, G., {et~al.} 2007, \apj, 670, 156, \dodoi{10.1086/521818}

\bibitem[{{D{\'\i}az-Garc{\'\i}a} {et~al.}(2019){D{\'\i}az-Garc{\'\i}a}, {Cenarro}, {L{\'o}pez-Sanjuan}, {Ferreras}, {Cervi{\~n}o}, {Fern{\'a}ndez-Soto}, {Gonz{\'a}lez Delgado}, {M{\'a}rquez}, {Povi{\'c}}, {San Roman}, {Viironen}, {Moles}, {Crist{\'o}bal-Hornillos}, {L{\'o}pez-Comazzi}, {Alfaro}, {Aparicio-Villegas}, {Ben{\'\i}tez}, {Broadhurst}, {Cabrera-Ca{\~n}o}, {Castander}, {Cepa}, {Husillos}, {Infante}, {Aguerri}, {Mart{\'\i}nez}, {Masegosa}, {Molino}, {del Olmo}, {Perea}, {Prada}, \& {Quintana}}]{diazgarcia2019}
{D{\'\i}az-Garc{\'\i}a}, L.~A., {Cenarro}, A.~J., {L{\'o}pez-Sanjuan}, C., {et~al.} 2019, \aap, 631, A156, \dodoi{10.1051/0004-6361/201832788}

\bibitem[{{Eales} {et~al.}(2017){Eales}, {de Vis}, {Smith}, {Appah}, {Ciesla}, {Duffield}, \& {Schofield}}]{eales2017}
{Eales}, S., {de Vis}, P., {Smith}, M. W.~L., {et~al.} 2017, \mnras, 465, 3125, \dodoi{10.1093/mnras/stw2875}

\bibitem[{{Elbaz} {et~al.}(2011){Elbaz}, {Dickinson}, {Hwang}, {D{\'\i}az-Santos}, {Magdis}, {Magnelli}, {Le Borgne}, {Galliano}, {Pannella}, {Chanial}, {Armus}, {Charmandaris}, {Daddi}, {Aussel}, {Popesso}, {Kartaltepe}, {Altieri}, {Valtchanov}, {Coia}, {Dannerbauer}, {Dasyra}, {Leiton}, {Mazzarella}, {Alexander}, {Buat}, {Burgarella}, {Chary}, {Gilli}, {Ivison}, {Juneau}, {Le Floc'h}, {Lutz}, {Morrison}, {Mullaney}, {Murphy}, {Pope}, {Scott}, {Brodwin}, {Calzetti}, {Cesarsky}, {Charlot}, {Dole}, {Eisenhardt}, {Ferguson}, {F{\"o}rster Schreiber}, {Frayer}, {Giavalisco}, {Huynh}, {Koekemoer}, {Papovich}, {Reddy}, {Surace}, {Teplitz}, {Yun}, \& {Wilson}}]{elbaz2011}
{Elbaz}, D., {Dickinson}, M., {Hwang}, H.~S., {et~al.} 2011, \aap, 533, A119, \dodoi{10.1051/0004-6361/201117239}

\bibitem[{{Feldmann}(2017)}]{feldman2017}
{Feldmann}, R. 2017, \mnras, 470, L59, \dodoi{10.1093/mnrasl/slx073}

\bibitem[{{Finlator} {et~al.}(2011){Finlator}, {Oppenheimer}, \& {Dav{\'e}}}]{finlator2011}
{Finlator}, K., {Oppenheimer}, B.~D., \& {Dav{\'e}}, R. 2011, \mnras, 410, 1703, \dodoi{10.1111/j.1365-2966.2010.17554.x}

\bibitem[{{Fitzpatrick} \& {Massa}(2007)}]{fitzpatrickmassa2007}
{Fitzpatrick}, E.~L., \& {Massa}, D. 2007, \apj, 663, 320, \dodoi{10.1086/518158}

\bibitem[{{Fontana} {et~al.}(2009){Fontana}, {Santini}, {Grazian}, {Pentericci}, {Fiore}, {Castellano}, {Giallongo}, {Menci}, {Salimbeni}, {Cristiani}, {Nonino}, \& {Vanzella}}]{fontana2009}
{Fontana}, A., {Santini}, P., {Grazian}, A., {et~al.} 2009, \aap, 501, 15, \dodoi{10.1051/0004-6361/200911650}

\bibitem[{Foreman-Mackey {et~al.}(2014)Foreman-Mackey, Sick, \& Johnson}]{python-FSPS}
Foreman-Mackey, D., Sick, J., \& Johnson, B. 2014, python-fsps: Python bindings to FSPS (v0.1.1), \dodoi{10.5281/zenodo.12157}

\bibitem[{{Gim{\'e}nez-Arteaga} {et~al.}(2022){Gim{\'e}nez-Arteaga}, {Brammer}, {Marchesini}, {Colina}, {Bajaj}, {Brinch}, {Calzetti}, {Lange-Vagle}, {Murphy}, {Perna}, {Piqueras-L{\'o}pez}, \& {Snyder}}]{gimenez2022}
{Gim{\'e}nez-Arteaga}, C., {Brammer}, G.~B., {Marchesini}, D., {et~al.} 2022, \apjs, 263, 17, \dodoi{10.3847/1538-4365/ac958c}

\bibitem[{{Gould} {et~al.}(2023){Gould}, {Brammer}, {Valentino}, {Whitaker}, {Weaver}, {Lagos}, {Rizzo}, {Franco}, {Hsieh}, {Ilbert}, {Jin}, {Magdis}, {McCracken}, {Mobasher}, {Shuntov}, {Steinhardt}, {Strait}, \& {Toft}}]{gould2023}
{Gould}, K. M.~L., {Brammer}, G., {Valentino}, F., {et~al.} 2023, \aj, 165, 248, \dodoi{10.3847/1538-3881/accadc}

\bibitem[{{Hao} {et~al.}(2011){Hao}, {Kennicutt}, {Johnson}, {Calzetti}, {Dale}, \& {Moustakas}}]{hao2011}
{Hao}, C.-N., {Kennicutt}, R.~C., {Johnson}, B.~D., {et~al.} 2011, \apj, 741, 124, \dodoi{10.1088/0004-637X/741/2/124}

\bibitem[{Harris {et~al.}(2020)Harris, Millman, van~der Walt, Gommers, Virtanen, Cournapeau, Wieser, Taylor, Berg, Smith, Kern, Picus, Hoyer, van Kerkwijk, Brett, Haldane, del R{\'{i}}o, Wiebe, Peterson, G{\'{e}}rard-Marchant, Sheppard, Reddy, Weckesser, Abbasi, Gohlke, \& Oliphant}]{numpy}
Harris, C.~R., Millman, K.~J., van~der Walt, S.~J., {et~al.} 2020, Nature, 585, 357, \dodoi{10.1038/s41586-020-2649-2}

\bibitem[{{Hogg} {et~al.}(2010){Hogg}, {Bovy}, \& {Lang}}]{hogg2010}
{Hogg}, D.~W., {Bovy}, J., \& {Lang}, D. 2010, arXiv e-prints, arXiv:1008.4686, \dodoi{10.48550/arXiv.1008.4686}

\bibitem[{Hunter(2007)}]{matplotlib}
Hunter, J.~D. 2007, Computing In Science \& Engineering, 9, 90, \dodoi{10.1109/MCSE.2007.55}

\bibitem[{{Johnson} {et~al.}(2021){Johnson}, {Leja}, {Conroy}, \& {Speagle}}]{johnson2021}
{Johnson}, B.~D., {Leja}, J., {Conroy}, C., \& {Speagle}, J.~S. 2021, \apjs, 254, 22, \dodoi{10.3847/1538-4365/abef67}

\bibitem[{{Kashino} {et~al.}(2023){Kashino}, {Lilly}, {Matthee}, {Eilers}, {Mackenzie}, {Bordoloi}, \& {Simcoe}}]{kashino2023}
{Kashino}, D., {Lilly}, S.~J., {Matthee}, J., {et~al.} 2023, \apj, 950, 66, \dodoi{10.3847/1538-4357/acc588}

\bibitem[{{Kennicutt}(1998)}]{kennicut1998}
{Kennicutt}, Robert~C., J. 1998, \araa, 36, 189, \dodoi{10.1146/annurev.astro.36.1.189}

\bibitem[{{Kennicutt} \& {Evans}(2012)}]{kennicutt2012}
{Kennicutt}, R.~C., \& {Evans}, N.~J. 2012, \araa, 50, 531, \dodoi{10.1146/annurev-astro-081811-125610}

\bibitem[{Kluyver {et~al.}(2016)Kluyver, Ragan-Kelley, P{\'e}rez, Granger, Bussonnier, Frederic, Kelley, Hamrick, Grout, Corlay, Ivanov, Avila, Abdalla, \& Willing}]{jupyter}
Kluyver, T., Ragan-Kelley, B., P{\'e}rez, F., {et~al.} 2016, in Positioning and Power in Academic Publishing: Players, Agents and Agendas, ed. F.~Loizides \& B.~Schmidt, IOS Press, 87 -- 90

\bibitem[{{Kokorev} {et~al.}(2022){Kokorev}, {Brammer}, {Fujimoto}, {Kohno}, {Magdis}, {Valentino}, {Toft}, {Oesch}, {Davidzon}, {Bauer}, {Coe}, {Egami}, {Oguri}, {Ouchi}, {Postman}, {Richard}, {Jolly}, {Knudsen}, {Sun}, {Weaver}, {Ao}, {Baker}, {Bradley}, {Caputi}, {Dessauges-Zavadsky}, {Espada}, {Hatsukade}, {Koekemoer}, {Mu{\~n}oz Arancibia}, {Shimasaku}, {Umehata}, {Wang}, \& {Wang}}]{kokorev2022}
{Kokorev}, V., {Brammer}, G., {Fujimoto}, S., {et~al.} 2022, \apjs, 263, 38, \dodoi{10.3847/1538-4365/ac9909}

\bibitem[{{Kriek} {et~al.}(2009){Kriek}, {van Dokkum}, {Labb{\'e}}, {Franx}, {Illingworth}, {Marchesini}, \& {Quadri}}]{kriek2009}
{Kriek}, M., {van Dokkum}, P.~G., {Labb{\'e}}, I., {et~al.} 2009, \apj, 700, 221, \dodoi{10.1088/0004-637X/700/1/221}

\bibitem[{{Kriek} {et~al.}(2010){Kriek}, {Labb{\'e}}, {Conroy}, {Whitaker}, {van Dokkum}, {Brammer}, {Franx}, {Illingworth}, {Marchesini}, {Muzzin}, {Quadri}, \& {Rudnick}}]{kriek2010}
{Kriek}, M., {Labb{\'e}}, I., {Conroy}, C., {et~al.} 2010, \apjl, 722, L64, \dodoi{10.1088/2041-8205/722/1/L64}

\bibitem[{{Kron}(1980)}]{kron1980}
{Kron}, R.~G. 1980, \apjs, 43, 305, \dodoi{10.1086/190669}

\bibitem[{{Kroupa}(2001)}]{kroupa2001}
{Kroupa}, P. 2001, \mnras, 322, 231, \dodoi{10.1046/j.1365-8711.2001.04022.x}

\bibitem[{{Kurczynski} {et~al.}(2016){Kurczynski}, {Gawiser}, {Acquaviva}, {Bell}, {Dekel}, {de Mello}, {Ferguson}, {Gardner}, {Grogin}, {Guo}, {Hopkins}, {Koekemoer}, {Koo}, {Lee}, {Mobasher}, {Primack}, {Rafelski}, {Soto}, \& {Teplitz}}]{Kurczynski2016}
{Kurczynski}, P., {Gawiser}, E., {Acquaviva}, V., {et~al.} 2016, \apjl, 820, L1, \dodoi{10.3847/2041-8205/820/1/L1}

\bibitem[{{Labb{\'e}} {et~al.}(2005){Labb{\'e}}, {Huang}, {Franx}, {Rudnick}, {Barmby}, {Daddi}, {van Dokkum}, {Fazio}, {F{\"o}rster Schreiber}, {Moorwood}, {Rix}, {R{\"o}ttgering}, {Trujillo}, \& {van der Werf}}]{labbe2005}
{Labb{\'e}}, I., {Huang}, J., {Franx}, M., {et~al.} 2005, \apjl, 624, L81, \dodoi{10.1086/430700}

\bibitem[{{Labb{\'e}} {et~al.}(2023){Labb{\'e}}, {van Dokkum}, {Nelson}, {Bezanson}, {Suess}, {Leja}, {Brammer}, {Whitaker}, {Mathews}, {Stefanon}, \& {Wang}}]{labbe2023}
{Labb{\'e}}, I., {van Dokkum}, P., {Nelson}, E., {et~al.} 2023, \nat, 616, 266, \dodoi{10.1038/s41586-023-05786-2}

\bibitem[{{Lee} {et~al.}(2015){Lee}, {Sanders}, {Casey}, {Toft}, {Scoville}, {Hung}, {Le Floc'h}, {Ilbert}, {Zahid}, {Aussel}, {Capak}, {Kartaltepe}, {Kewley}, {Li}, {Schawinski}, {Sheth}, \& {Xiao}}]{lee2015}
{Lee}, N., {Sanders}, D.~B., {Casey}, C.~M., {et~al.} 2015, \apj, 801, 80, \dodoi{10.1088/0004-637X/801/2/80}

\bibitem[{{Leja} {et~al.}(2017){Leja}, {Johnson}, {Conroy}, {van Dokkum}, \& {Byler}}]{leja2017}
{Leja}, J., {Johnson}, B.~D., {Conroy}, C., {van Dokkum}, P.~G., \& {Byler}, N. 2017, \apj, 837, 170, \dodoi{10.3847/1538-4357/aa5ffe}

\bibitem[{{Leja} {et~al.}(2019){Leja}, {Tacchella}, \& {Conroy}}]{leja2019}
{Leja}, J., {Tacchella}, S., \& {Conroy}, C. 2019, \apjl, 880, L9, \dodoi{10.3847/2041-8213/ab2f8c}

\bibitem[{{Leja} {et~al.}(2022){Leja}, {Speagle}, {Ting}, {Johnson}, {Conroy}, {Whitaker}, {Nelson}, {van Dokkum}, \& {Franx}}]{leja2022}
{Leja}, J., {Speagle}, J.~S., {Ting}, Y.-S., {et~al.} 2022, \apj, 936, 165, \dodoi{10.3847/1538-4357/ac887d}

\bibitem[{{Leslie} {et~al.}(2020){Leslie}, {Schinnerer}, {Liu}, {Magnelli}, {Algera}, {Karim}, {Davidzon}, {Gozaliasl}, {Jim{\'e}nez-Andrade}, {Lang}, {Sargent}, {Novak}, {Groves}, {Smol{\v{c}}i{\'c}}, {Zamorani}, {Vaccari}, {Battisti}, {Vardoulaki}, {Peng}, \& {Kartaltepe}}]{leslie2020}
{Leslie}, S.~K., {Schinnerer}, E., {Liu}, D., {et~al.} 2020, \apj, 899, 58, \dodoi{10.3847/1538-4357/aba044}

\bibitem[{{Lower} {et~al.}(2020){Lower}, {Narayanan}, {Leja}, {Johnson}, {Conroy}, \& {Dav{\'e}}}]{lower2020}
{Lower}, S., {Narayanan}, D., {Leja}, J., {et~al.} 2020, \apj, 904, 33, \dodoi{10.3847/1538-4357/abbfa7}

\bibitem[{{Madau} \& {Dickinson}(2014)}]{madau2014}
{Madau}, P., \& {Dickinson}, M. 2014, \araa, 52, 415, \dodoi{10.1146/annurev-astro-081811-125615}

\bibitem[{{M{\'e}rida} {et~al.}(2023){M{\'e}rida}, {P{\'e}rez-Gonz{\'a}lez}, {S{\'a}nchez-Bl{\'a}zquez}, {Garc{\'\i}a-Argum{\'a}nez}, {Annunziatella}, {Costantin}, {Lumbreras-Calle}, {Alcalde-Pampliega}, {Barro}, {Espino-Briones}, \& {Koekemoer}}]{merida2023}
{M{\'e}rida}, R.~M., {P{\'e}rez-Gonz{\'a}lez}, P.~G., {S{\'a}nchez-Bl{\'a}zquez}, P., {et~al.} 2023, arXiv e-prints, arXiv:2303.16234, \dodoi{10.48550/arXiv.2303.16234}

\bibitem[{{Merlin} {et~al.}(2018){Merlin}, {Fontana}, {Castellano}, {Santini}, {Torelli}, {Boutsia}, {Wang}, {Grazian}, {Pentericci}, {Schreiber}, {Ciesla}, {McLure}, {Derriere}, {Dunlop}, \& {Elbaz}}]{merlin2018}
{Merlin}, E., {Fontana}, A., {Castellano}, M., {et~al.} 2018, \mnras, 473, 2098, \dodoi{10.1093/mnras/stx2385}

\bibitem[{{Momcheva} {et~al.}(2016){Momcheva}, {Brammer}, {van Dokkum}, {Skelton}, {Whitaker}, {Nelson}, {Fumagalli}, {Maseda}, {Leja}, {Franx}, {Rix}, {Bezanson}, {Da Cunha}, {Dickey}, {F{\"o}rster Schreiber}, {Illingworth}, {Kriek}, {Labb{\'e}}, {Ulf Lange}, {Lundgren}, {Magee}, {Marchesini}, {Oesch}, {Pacifici}, {Patel}, {Price}, {Tal}, {Wake}, {van der Wel}, \& {Wuyts}}]{momcheva2016}
{Momcheva}, I.~G., {Brammer}, G.~B., {van Dokkum}, P.~G., {et~al.} 2016, \apjs, 225, 27, \dodoi{10.3847/0067-0049/225/2/27}

\bibitem[{{Mowla} {et~al.}(2019){Mowla}, {van Dokkum}, {Brammer}, {Momcheva}, {van der Wel}, {Whitaker}, {Nelson}, {Bezanson}, {Muzzin}, {Franx}, {MacKenty}, {Leja}, {Kriek}, \& {Marchesini}}]{mowla2019}
{Mowla}, L.~A., {van Dokkum}, P., {Brammer}, G.~B., {et~al.} 2019, \apj, 880, 57, \dodoi{10.3847/1538-4357/ab290a}

\bibitem[{{Murphy} {et~al.}(2011){Murphy}, {Condon}, {Schinnerer}, {Kennicutt}, {Calzetti}, {Armus}, {Helou}, {Turner}, {Aniano}, {Beir{\~a}o}, {Bolatto}, {Brandl}, {Croxall}, {Dale}, {Donovan Meyer}, {Draine}, {Engelbracht}, {Hunt}, {Hao}, {Koda}, {Roussel}, {Skibba}, \& {Smith}}]{murphy2011}
{Murphy}, E.~J., {Condon}, J.~J., {Schinnerer}, E., {et~al.} 2011, \apj, 737, 67, \dodoi{10.1088/0004-637X/737/2/67}

\bibitem[{{Muzzin} {et~al.}(2013){Muzzin}, {Marchesini}, {Stefanon}, {Franx}, {McCracken}, {Milvang-Jensen}, {Dunlop}, {Fynbo}, {Brammer}, {Labb{\'e}}, \& {van Dokkum}}]{muzzin2013}
{Muzzin}, A., {Marchesini}, D., {Stefanon}, M., {et~al.} 2013, \apj, 777, 18, \dodoi{10.1088/0004-637X/777/1/18}

\bibitem[{{Nagaraj} {et~al.}(2021){Nagaraj}, {Ciardullo}, {Lawson}, {Bowman}, {Zeimann}, {Yang}, \& {Gronwall}}]{nagaraj2021}
{Nagaraj}, G., {Ciardullo}, R., {Lawson}, A., {et~al.} 2021, \apj, 912, 145, \dodoi{10.3847/1538-4357/abefcf}

\bibitem[{{Nelson} {et~al.}(2021){Nelson}, {Tacchella}, {Diemer}, {Leja}, {Hernquist}, {Whitaker}, {Weinberger}, {Pillepich}, {Nelson}, {Terrazas}, {Nevin}, {Brammer}, {Burkhart}, {Cochrane}, {van Dokkum}, {Johnson}, {Marinacci}, {Mowla}, {Pakmor}, {Skelton}, {Speagle}, {Springel}, {Torrey}, {Vogelsberger}, \& {Wuyts}}]{nelson2021}
{Nelson}, E.~J., {Tacchella}, S., {Diemer}, B., {et~al.} 2021, \mnras, 508, 219, \dodoi{10.1093/mnras/stab2131}

\bibitem[{{Noeske} {et~al.}(2007){Noeske}, {Weiner}, {Faber}, {Papovich}, {Koo}, {Somerville}, {Bundy}, {Conselice}, {Newman}, {Schiminovich}, {Le Floc'h}, {Coil}, {Rieke}, {Lotz}, {Primack}, {Barmby}, {Cooper}, {Davis}, {Ellis}, {Fazio}, {Guhathakurta}, {Huang}, {Kassin}, {Martin}, {Phillips}, {Rich}, {Small}, {Willmer}, \& {Wilson}}]{noeske2007}
{Noeske}, K.~G., {Weiner}, B.~J., {Faber}, S.~M., {et~al.} 2007, \apjl, 660, L43, \dodoi{10.1086/517926}

\bibitem[{{Oesch} {et~al.}(2023){Oesch}, {Brammer}, {Naidu}, {Bouwens}, {Chisholm}, {Illingworth}, {Matthee}, {Nelson}, {Qin}, {Reddy}, {Shapley}, {Shivaei}, {van Dokkum}, {Weibel}, {Whitaker}, {Wuyts}, {Covelo-Paz}, {Endsley}, {Fudamoto}, {Giovinazzo}, {Herard-Demanche}, {Kerutt}, {Kramarenko}, {Labbe}, {Leonova}, {Lin}, {Magee}, {Marchesini}, {Maseda}, {Mason}, {Matharu}, {Meyer}, {Neufeld}, {Prieto Lyon}, {Schaerer}, {Sharma}, {Shuntov}, {Smit}, {Stefanon}, {Wyithe}, \& {Xiao}}]{oesch23}
{Oesch}, P.~A., {Brammer}, G., {Naidu}, R.~P., {et~al.} 2023, arXiv e-prints, arXiv:2304.02026, \dodoi{10.48550/arXiv.2304.02026}

\bibitem[{{Oke} \& {Gunn}(1983)}]{okegunn}
{Oke}, J.~B., \& {Gunn}, J.~E. 1983, \apj, 266, 713, \dodoi{10.1086/160817}

\bibitem[{{Osterbrock}(1989)}]{osterbrock1998}
{Osterbrock}, D.~E. 1989, {Astrophysics of gaseous nebulae and active galactic nuclei}

\bibitem[{{Pacifici} {et~al.}(2023){Pacifici}, {Iyer}, {Mobasher}, {da Cunha}, {Acquaviva}, {Burgarella}, {Calistro Rivera}, {Carnall}, {Chang}, {Chartab}, {Cooke}, {Fairhurst}, {Kartaltepe}, {Leja}, {Ma{\l}ek}, {Salmon}, {Torelli}, {Vidal-Garc{\'\i}a}, {Boquien}, {Brammer}, {Brown}, {Capak}, {Chevallard}, {Circosta}, {Croton}, {Davidzon}, {Dickinson}, {Duncan}, {Faber}, {Ferguson}, {Fontana}, {Guo}, {Haeussler}, {Hemmati}, {Jafariyazani}, {Kassin}, {Larson}, {Lee}, {Mantha}, {Marchi}, {Nayyeri}, {Newman}, {Pandya}, {Pforr}, {Reddy}, {Sanders}, {Shah}, {Shahidi}, {Stevans}, {Triani}, {Tyler}, {Vanderhoof}, {de la Vega}, {Wang}, \& {Weston}}]{pacifici2023}
{Pacifici}, C., {Iyer}, K.~G., {Mobasher}, B., {et~al.} 2023, \apj, 944, 141, \dodoi{10.3847/1538-4357/acacff}

\bibitem[{{Papovich} {et~al.}(2009){Papovich}, {Rudnick}, {Rigby}, {Willmer}, {Smith}, {Finkelstein}, {Egami}, \& {Rieke}}]{papovich2009}
{Papovich}, C., {Rudnick}, G., {Rigby}, J.~R., {et~al.} 2009, \apj, 704, 1506, \dodoi{10.1088/0004-637X/704/2/1506}

\bibitem[{{Pasha} {et~al.}(2020){Pasha}, {Leja}, {van Dokkum}, {Conroy}, \& {Johnson}}]{pasha2020}
{Pasha}, I., {Leja}, J., {van Dokkum}, P.~G., {Conroy}, C., \& {Johnson}, B.~D. 2020, \apj, 898, 165, \dodoi{10.3847/1538-4357/aba0b1}

\bibitem[{{Patel} {et~al.}(2012){Patel}, {Holden}, {Kelson}, {Franx}, {van der Wel}, \& {Illingworth}}]{patel2012}
{Patel}, S.~G., {Holden}, B.~P., {Kelson}, D.~D., {et~al.} 2012, \apjl, 748, L27, \dodoi{10.1088/2041-8205/748/2/L27}

\bibitem[{P\'erez \& Granger(2007)}]{ipython}
P\'erez, F., \& Granger, B.~E. 2007, Computing in Science and Engineering, 9, 21, \dodoi{10.1109/MCSE.2007.53}

\bibitem[{{Popesso} {et~al.}(2023){Popesso}, {Concas}, {Cresci}, {Belli}, {Rodighiero}, {Inami}, {Dickinson}, {Ilbert}, {Pannella}, \& {Elbaz}}]{popesso2023}
{Popesso}, P., {Concas}, A., {Cresci}, G., {et~al.} 2023, \mnras, 519, 1526, \dodoi{10.1093/mnras/stac3214}

\bibitem[{{Price} {et~al.}(2014){Price}, {Kriek}, {Brammer}, {Conroy}, {F{\"o}rster Schreiber}, {Franx}, {Fumagalli}, {Lundgren}, {Momcheva}, {Nelson}, {Skelton}, {van Dokkum}, {Whitaker}, \& {Wuyts}}]{price2014}
{Price}, S.~H., {Kriek}, M., {Brammer}, G.~B., {et~al.} 2014, \apj, 788, 86, \dodoi{10.1088/0004-637X/788/1/86}

\bibitem[{{Reddy} {et~al.}(2023){Reddy}, {Topping}, {Sanders}, {Shapley}, \& {Brammer}}]{reddy2023}
{Reddy}, N.~A., {Topping}, M.~W., {Sanders}, R.~L., {Shapley}, A.~E., \& {Brammer}, G. 2023, \apj, 948, 83, \dodoi{10.3847/1538-4357/acc869}

\bibitem[{{Reddy} {et~al.}(2015){Reddy}, {Kriek}, {Shapley}, {Freeman}, {Siana}, {Coil}, {Mobasher}, {Price}, {Sanders}, \& {Shivaei}}]{reddy2015}
{Reddy}, N.~A., {Kriek}, M., {Shapley}, A.~E., {et~al.} 2015, \apj, 806, 259, \dodoi{10.1088/0004-637X/806/2/259}

\bibitem[{{Renzini}(2009)}]{renzini2009}
{Renzini}, A. 2009, \mnras, 398, L58, \dodoi{10.1111/j.1745-3933.2009.00710.x}

\bibitem[{{Renzini} \& {Peng}(2015)}]{renzini2015}
{Renzini}, A., \& {Peng}, Y.-j. 2015, \apjl, 801, L29, \dodoi{10.1088/2041-8205/801/2/L29}

\bibitem[{{Rieke} {et~al.}(2009){Rieke}, {Alonso-Herrero}, {Weiner}, {P{\'e}rez-Gonz{\'a}lez}, {Blaylock}, {Donley}, \& {Marcillac}}]{rieke2009}
{Rieke}, G.~H., {Alonso-Herrero}, A., {Weiner}, B.~J., {et~al.} 2009, \apj, 692, 556, \dodoi{10.1088/0004-637X/692/1/556}

\bibitem[{{Salim} \& {Narayanan}(2020)}]{salim2020}
{Salim}, S., \& {Narayanan}, D. 2020, \araa, 58, 529, \dodoi{10.1146/annurev-astro-032620-021933}

\bibitem[{{Salmon} {et~al.}(2015){Salmon}, {Papovich}, {Finkelstein}, {Tilvi}, {Finlator}, {Behroozi}, {Dahlen}, {Dav{\'e}}, {Dekel}, {Dickinson}, {Ferguson}, {Giavalisco}, {Long}, {Lu}, {Mobasher}, {Reddy}, {Somerville}, \& {Wechsler}}]{salmon2015}
{Salmon}, B., {Papovich}, C., {Finkelstein}, S.~L., {et~al.} 2015, \apj, 799, 183, \dodoi{10.1088/0004-637X/799/2/183}

\bibitem[{Santini {et~al.}(2017)Santini, Fontana, Castellano, Criscienzo, Merlin, Amorin, Cullen, Daddi, Dickinson, Dunlop, Grazian, Lamastra, McLure, Micha{\l}owski, Pentericci, \& Shu}]{santini2017}
Santini, P., Fontana, A., Castellano, M., {et~al.} 2017, The Astrophysical Journal, 847, 76, \dodoi{10.3847/1538-4357/aa8874}

\bibitem[{{Schreiber} {et~al.}(2015){Schreiber}, {Pannella}, {Elbaz}, {B{\'e}thermin}, {Inami}, {Dickinson}, {Magnelli}, {Wang}, {Aussel}, {Daddi}, {Juneau}, {Shu}, {Sargent}, {Buat}, {Faber}, {Ferguson}, {Giavalisco}, {Koekemoer}, {Magdis}, {Morrison}, {Papovich}, {Santini}, \& {Scott}}]{schreiber2015}
{Schreiber}, C., {Pannella}, M., {Elbaz}, D., {et~al.} 2015, \aap, 575, A74, \dodoi{10.1051/0004-6361/201425017}

\bibitem[{{Sherman} {et~al.}(2020){Sherman}, {Jogee}, {Florez}, {Stevans}, {Kawinwanichakij}, {Wold}, {Finkelstein}, {Papovich}, {Acquaviva}, {Ciardullo}, {Gronwall}, \& {Escalante}}]{sherman2020}
{Sherman}, S., {Jogee}, S., {Florez}, J., {et~al.} 2020, \mnras, 491, 3318, \dodoi{10.1093/mnras/stz3229}

\bibitem[{{Sherman} {et~al.}(2021){Sherman}, {Jogee}, {Florez}, {Finkelstein}, {Ciardullo}, {Wold}, {Stevans}, {Kawinwanichakij}, {Papovich}, \& {Gronwall}}]{sherman2021}
---. 2021, \mnras, 505, 947, \dodoi{10.1093/mnras/stab1350}

\bibitem[{{Shivaei} {et~al.}(2015){Shivaei}, {Reddy}, {Shapley}, {Kriek}, {Siana}, {Mobasher}, {Coil}, {Freeman}, {Sanders}, {Price}, {de Groot}, \& {Azadi}}]{shivaei2015}
{Shivaei}, I., {Reddy}, N.~A., {Shapley}, A.~E., {et~al.} 2015, \apj, 815, 98, \dodoi{10.1088/0004-637X/815/2/98}

\bibitem[{{Shivaei} {et~al.}(2017){Shivaei}, {Reddy}, {Shapley}, {Siana}, {Kriek}, {Mobasher}, {Coil}, {Freeman}, {Sanders}, {Price}, {Azadi}, \& {Zick}}]{shivaei2017}
---. 2017, \apj, 837, 157, \dodoi{10.3847/1538-4357/aa619c}

\bibitem[{{Shivaei} {et~al.}(2020){Shivaei}, {Reddy}, {Rieke}, {Shapley}, {Kriek}, {Battisti}, {Mobasher}, {Sanders}, {Fetherolf}, {Azadi}, {Coil}, {Freeman}, {de Groot}, {Leung}, {Price}, {Siana}, \& {Zick}}]{shivaei2020}
{Shivaei}, I., {Reddy}, N., {Rieke}, G., {et~al.} 2020, \apj, 899, 117, \dodoi{10.3847/1538-4357/aba35e}

\bibitem[{{Skelton} {et~al.}(2014){Skelton}, {Whitaker}, {Momcheva}, {Brammer}, {van Dokkum}, {Labb{\'e}}, {Franx}, {van der Wel}, {Bezanson}, {Da Cunha}, {Fumagalli}, {F{\"o}rster Schreiber}, {Kriek}, {Leja}, {Lundgren}, {Magee}, {Marchesini}, {Maseda}, {Nelson}, {Oesch}, {Pacifici}, {Patel}, {Price}, {Rix}, {Tal}, {Wake}, \& {Wuyts}}]{skelton2014}
{Skelton}, R.~E., {Whitaker}, K.~E., {Momcheva}, I.~G., {et~al.} 2014, \apjs, 214, 24, \dodoi{10.1088/0067-0049/214/2/24}

\bibitem[{{Somerville} \& {Dav{\'e}}(2015)}]{somervilladave2015}
{Somerville}, R.~S., \& {Dav{\'e}}, R. 2015, \araa, 53, 51, \dodoi{10.1146/annurev-astro-082812-140951}

\bibitem[{Speagle(2018)}]{dynesty}
Speagle, J. 2018, dynesty, \url{https://github.com/joshspeagle/dynesty},  GitHub

\bibitem[{{Speagle} {et~al.}(2014){Speagle}, {Steinhardt}, {Capak}, \& {Silverman}}]{speagle2014}
{Speagle}, J.~S., {Steinhardt}, C.~L., {Capak}, P.~L., \& {Silverman}, J.~D. 2014, \apjs, 214, 15, \dodoi{10.1088/0067-0049/214/2/15}

\bibitem[{{Thomas} {et~al.}(2005){Thomas}, {Maraston}, {Bender}, \& {Mendes de Oliveira}}]{thomas2005}
{Thomas}, D., {Maraston}, C., {Bender}, R., \& {Mendes de Oliveira}, C. 2005, \apj, 621, 673, \dodoi{10.1086/426932}

\bibitem[{{Tomczak} {et~al.}(2016){Tomczak}, {Quadri}, {Tran}, {Labb{\'e}}, {Straatman}, {Papovich}, {Glazebrook}, {Allen}, {Brammer}, {Cowley}, {Dickinson}, {Elbaz}, {Inami}, {Kacprzak}, {Morrison}, {Nanayakkara}, {Persson}, {Rees}, {Salmon}, {Schreiber}, {Spitler}, \& {Whitaker}}]{tomczak2016}
{Tomczak}, A.~R., {Quadri}, R.~F., {Tran}, K.-V.~H., {et~al.} 2016, \apj, 817, 118, \dodoi{10.3847/0004-637X/817/2/118}

\bibitem[{{Virtanen} {et~al.}(2020){Virtanen}, {Gommers}, {Oliphant}, {Haberland}, {Reddy}, {Cournapeau}, {Burovski}, {Peterson}, {Weckesser}, {Bright}, {van der Walt}, {Brett}, {Wilson}, {Millman}, {Mayorov}, {Nelson}, {Jones}, {Kern}, {Larson}, {Carey}, {Polat}, {Feng}, {Moore}, {VanderPlas}, {Laxalde}, {Perktold}, {Cimrman}, {Henriksen}, {Quintero}, {Harris}, {Archibald}, {Ribeiro}, {Pedregosa}, {van Mulbregt}, \& {SciPy 1. 0 Contributors}}]{scipy}
{Virtanen}, P., {Gommers}, R., {Oliphant}, T.~E., {et~al.} 2020, Nature Methods, 17, 261, \dodoi{10.1038/s41592-019-0686-2}

\bibitem[{Waskom {et~al.}(2018)Waskom, Botvinnik, O'Kane, Hobson, Ostblom, Lukauskas, Gemperline, Augspurger, Halchenko, Cole, Warmenhoven, de~Ruiter, Pye, Hoyer, Vanderplas, Villalba, Kunter, Quintero, Bachant, Martin, Meyer, Miles, Ram, Brunner, Yarkoni, Williams, Evans, Fitzgerald, Brian, \& Qalieh}]{seaborn}
Waskom, M., Botvinnik, O., O'Kane, D., {et~al.} 2018, mwaskom/seaborn: v0.9.0 (July 2018), \dodoi{10.5281/zenodo.1313201}

\bibitem[{{Wen} {et~al.}(2013){Wen}, {Wu}, {Zhu}, {Lam}, {Wu}, {Wicker}, \& {Zhao}}]{wen2013}
{Wen}, X.-Q., {Wu}, H., {Zhu}, Y.-N., {et~al.} 2013, \mnras, 433, 2946, \dodoi{10.1093/mnras/stt939}

\bibitem[{{Whitaker} {et~al.}(2012{\natexlab{a}}){Whitaker}, {Kriek}, {van Dokkum}, {Bezanson}, {Brammer}, {Franx}, \& {Labb{\'e}}}]{whitaker2012b}
{Whitaker}, K.~E., {Kriek}, M., {van Dokkum}, P.~G., {et~al.} 2012{\natexlab{a}}, \apj, 745, 179, \dodoi{10.1088/0004-637X/745/2/179}

\bibitem[{{Whitaker} {et~al.}(2012{\natexlab{b}}){Whitaker}, {van Dokkum}, {Brammer}, \& {Franx}}]{whitaker2012}
{Whitaker}, K.~E., {van Dokkum}, P.~G., {Brammer}, G., \& {Franx}, M. 2012{\natexlab{b}}, \apjl, 754, L29, \dodoi{10.1088/2041-8205/754/2/L29}

\bibitem[{{Whitaker} {et~al.}(2011){Whitaker}, {Labb{\'e}}, {van Dokkum}, {Brammer}, {Kriek}, {Marchesini}, {Quadri}, {Franx}, {Muzzin}, {Williams}, {Bezanson}, {Illingworth}, {Lee}, {Lundgren}, {Nelson}, {Rudnick}, {Tal}, \& {Wake}}]{whitaker2011}
{Whitaker}, K.~E., {Labb{\'e}}, I., {van Dokkum}, P.~G., {et~al.} 2011, \apj, 735, 86, \dodoi{10.1088/0004-637X/735/2/86}

\bibitem[{{Whitaker} {et~al.}(2014){Whitaker}, {Franx}, {Leja}, {van Dokkum}, {Henry}, {Skelton}, {Fumagalli}, {Momcheva}, {Brammer}, {Labb{\'e}}, {Nelson}, \& {Rigby}}]{whitaker2014}
{Whitaker}, K.~E., {Franx}, M., {Leja}, J., {et~al.} 2014, \apj, 795, 104, \dodoi{10.1088/0004-637X/795/2/104}

\bibitem[{{Whitaker} {et~al.}(2015){Whitaker}, {Franx}, {Bezanson}, {Brammer}, {van Dokkum}, {Kriek}, {Labb{\'e}}, {Leja}, {Momcheva}, {Nelson}, {Rigby}, {Rix}, {Skelton}, {van der Wel}, \& {Wuyts}}]{whitaker2015}
{Whitaker}, K.~E., {Franx}, M., {Bezanson}, R., {et~al.} 2015, \apjl, 811, L12, \dodoi{10.1088/2041-8205/811/1/L12}

\bibitem[{{Williams} {et~al.}(2009){Williams}, {Quadri}, {Franx}, {van Dokkum}, \& {Labb{\'e}}}]{williams2009}
{Williams}, R.~J., {Quadri}, R.~F., {Franx}, M., {van Dokkum}, P., \& {Labb{\'e}}, I. 2009, \apj, 691, 1879, \dodoi{10.1088/0004-637X/691/2/1879}

\bibitem[{{Wuyts} {et~al.}(2007){Wuyts}, {Labb{\'e}}, {Franx}, {Rudnick}, {van Dokkum}, {Fazio}, {F{\"o}rster Schreiber}, {Huang}, {Moorwood}, {Rix}, {R{\"o}ttgering}, \& {van der Werf}}]{wuyts2007}
{Wuyts}, S., {Labb{\'e}}, I., {Franx}, M., {et~al.} 2007, \apj, 655, 51, \dodoi{10.1086/509708}

\end{thebibliography}
\bibliographystyle{aasjournal}

\end{document}